\documentclass[a4paper,fleqn]{cas-dc}

\usepackage{amsthm}
\usepackage{tikz}
\usepackage{tcolorbox}
\usepackage{subcaption}
\usepackage[linesnumbered, vlined, ruled,algo2e]{algorithm2e}
\usepackage{xcolor,colortbl}
\usepackage[dvipsnames]{xcolor}

 \usepackage{tabularray}
 \usepackage{amsmath}
 \usepackage{amssymb}
 \usepackage{needspace}
 \usepackage{booktabs}
 \usepackage[inline]{enumitem}

\newcommand{\tabsection}[1]{%
  \addlinespace[6pt]
  \multicolumn{2}{@{}l}{\textbf{\textsc{#1}}} \\
  \ibline
}
\usepackage{makecell}

\definecolor{dark-gray}{gray}{0.2}

\newcommand{\eat}[1]{}
 
 \newcommand{\hidefornow}[1]{}        

\newcommand\dunderline[3][-1pt]{{%
  \sbox0{#3}%
  \ooalign{\copy0\cr\rule[\dimexpr#1-#2\relax]{\wd0}{#2}}}}

\newcommand{\stitle}[1]{\vspace{14pt}\noindent\textbf{#1}} 
\newcommand{\sititle}[1]{\bigskip\noindent\dunderline[-2pt]{0.5pt}{\textit{#1}}}

\newcommand{\fnc}[1]{\scalebox{0.9}{\textsf{#1}}}

\DeclareMathOperator*{\argmax}{arg\,max}

\DeclareMathOperator*{\mymin}{min}

\newcounter{ex}

\newcommand{\myFontP}[1]{{\fontfamily{ppl}\selectfont #1}} 

\DontPrintSemicolon
\SetAlgoCaptionLayout{small}
\SetProcNameSty{small}
\SetProcArgSty{small}
\SetAlgoInsideSkip{}
\SetAlgoSkip{medskip}
\SetVlineSkip {1.5mm}
\SetAlgoCaptionSeparator{.}

\SetKw{Logand}{and}
\SetKw{Logor}{or}
\SetKw{Lognot}{not}
\SetKw{mybreak}{break}
\SetKw{mycontinue}{continue}

\SetKwInput{KwVar}{Variables}
\SetKwInput{KwPar}{Parameters}
\SetKwRepeat{Do}{do}{while}

\newcommand*\circlednum[1]{\tikz[baseline=(char.base)]{
		\node[shape=circle,draw,inner sep=0.75pt, line width=1.0pt,text=black](char) {\footnotesize{\textbf{#1}}};}}

\newcommand{\mycomment} [1] 	{\tiny{\textcolor{dark-gray}{\myFontP{{#1}}}}}
\SetKwComment{Comment}{\mycomment{/\!\!/}}{}

\SetKwBlock{myblock}{beginnikos}{end} 

\SetKwData{myfalse}{false}
\SetKwData{mytrue}{true}

\usepackage{tabularx}

\newcommand{\tline}  {\specialrule{0.8 pt}{0pt}{1pt}}		 
\newcommand{\bline}  {\specialrule{0.8 pt}{1pt}{0pt}}		
\newcommand{\dlineB}  {\specialrule{0.8 pt}{1pt}{2pt} \specialrule{0.8 pt}{0pt}{0pt}}		

\newcommand{\ibline}  {\specialrule{0.6 pt}{1pt}{1pt}} 		

\newcounter{corollcnt}

\graphicspath{{plots/}}

\usepackage[numbers]{natbib}

\def\tsc#1{\csdef{#1}{\textsc{\lowercase{#1}}\xspace}}
\tsc{WGM}
\tsc{QE}

\begin{document}
\let\WriteBookmarks\relax
\def\floatpagepagefraction{1}
\def\textpagefraction{.001}

\shorttitle{Explaining Group Recommendations via Counterfactuals}
\shortauthors{M. Stratigi, N. Bikakis, K. Stefanidis}

\title [mode = title]{Explaining Group Recommendations via Counterfactuals}

 \author[1]{Maria Stratigi}
  [orcid=0000-0003-2482-4605]
  \cormark[1]
\ead{maria.stratigi@tuni.fi}

  \author[2,3]{Nikos Bikakis}
  [orcid=0000-0001-6859-1941]
 \ead{bikakis@hmu.gr}

  \author[1]{Kostas Stefanidis}
  [orcid=0000-0003-1317-8062]
\ead{konstantinos.stefanidis@tuni.fi}

  \affiliation[1]{organization={Data Science Research Centre, Tampere University},
  	city={Tampere},
  	country={Finland}}
  	    \affiliation[2]{organization={Hellenic Mediterranean University},
  	city={Chania},
  	country={Greece}}
  \affiliation[3]{organization={Athena Research Center},
  	country={Greece}}

\cortext[1]{Corresponding author}

\begin{abstract}
  Group recommender systems help users make collective choices but often lack transparency, leaving group members uncertain about why items are suggested. Existing explanation methods focus on individuals, offering limited support for groups where multiple preferences interact.
In this work, we introduce and formalize the problem of finding counterfactual explanations for group recommendations.
We introduce a comprehensive set of item- and explanation-level metrics that capture different aspects, such as relevance, popularity, explanation utility, interpretability, and fairness.
We design five algorithms that employ techniques such as search-space pruning, greedy selection, early stopping, and incremental refinement to reduce computational cost and the number of recommender calls while producing high-quality explanations.
The algorithms consider different trade-offs; for example, some prioritize computational efficiency at the expense of explanation quality, while others focus on improving explanation quality at a higher computational cost.
Experiments on real-world datasets analyze these trade-offs in terms of explanation size, computational cost, utility, and fairness.
\end{abstract}


\maketitle

\hypersetup{
	colorlinks,%
	citecolor=black,%
	filecolor=black,
	linkcolor=black,%
	urlcolor=black,%
	pdftitle={Explaining Group Recommendations via Counterfactuals},
	pdfauthor={Maria Stratigi, Nikos Bikakis,  Kostas Stefanidis},
	pdfsubject={Explainable AI, Fairness in Recommendation, Group Fairness, Counterfactual Reasoning, Model-agnostic Explanations, Group Recommender Systems, Explainable Recommender Systems,  Explainability in Recommendations},
	pdfkeywords={Explainable AI, Fairness in Recommendation, Group Fairness, Counterfactual Reasoning, Model-agnostic Explanations, Group Recommender Systems, Explainable Recommender Systems, Explainability in Recommendations}
}

\section{Introduction}
Recommender systems have become indispensable in a wide range of online platforms, from e-commerce and streaming services to tourism and education. While most research in recommender systems has traditionally focused on individuals, many real-world decision-making processes involve groups of users, e.g., families choosing a movie, friends selecting a restaurant, or colleagues planning a trip. Group recommender systems aim to support such scenarios by aggregating individual preferences into a joint recommendation. Despite their growing importance, they often remain opaque: users typically see the final output but have little insight into why certain items were recommended to the group. This lack of transparency undermines user trust and limits acceptance, particularly in settings where group members need to justify or negotiate decisions.

Explainability has emerged as a critical requirement for recommender systems \cite{DBLP:conf/icde/TintarevM07}. Existing research on explainable AI has explored various forms of explanations, including feature-based justifications \cite{lundberg2017unified}, rule-based reasoning \cite{vilone2020comparative}, and, more recently, counterfactual explanations \cite{sokol2019counterfactual}. Counterfactuals answer “what if” questions by identifying the changes needed in the input that would alter a recommendation outcome. For instance, a system may explain the appearance of a movie in a user’s list by stating:
“\textit{If you had not rated item X, this recommendation would not appear}”.
Counterfactual explanations are actionable, model-agnostic, and easily interpretable, making them  suitable for increasing transparency in black-box systems. However, prior work has almost exclusively focused on individual users \cite{2020_WSDM_GBRW, 10.1145/3589335.3651484,10.1145/3450613.3456846,BarkanBGAK24}, leaving group settings largely unexplored.

Explaining group recommendations introduces unique challenges. Unlike individuals, groups combine multiple preference profiles, often leading to complex dynamics of influence and compromise. A counterfactual explanation must therefore account not only for the system’s internal logic but also for how group members’ interactions collectively shape the outcome.
Moreover, group explanations raise additional concerns of fairness: if an explanation suggests removing only one member’s interactions, that user may perceive the system as biased or exclusionary.
Balancing interpretability, utility, and fairness is thus essential for explanations to be both informative and acceptable in group contexts \cite{DBLP:conf/recsys/Tintarev07}.

\begin{figure*}[t]
     \centering
     \includegraphics[width=0.99\linewidth]{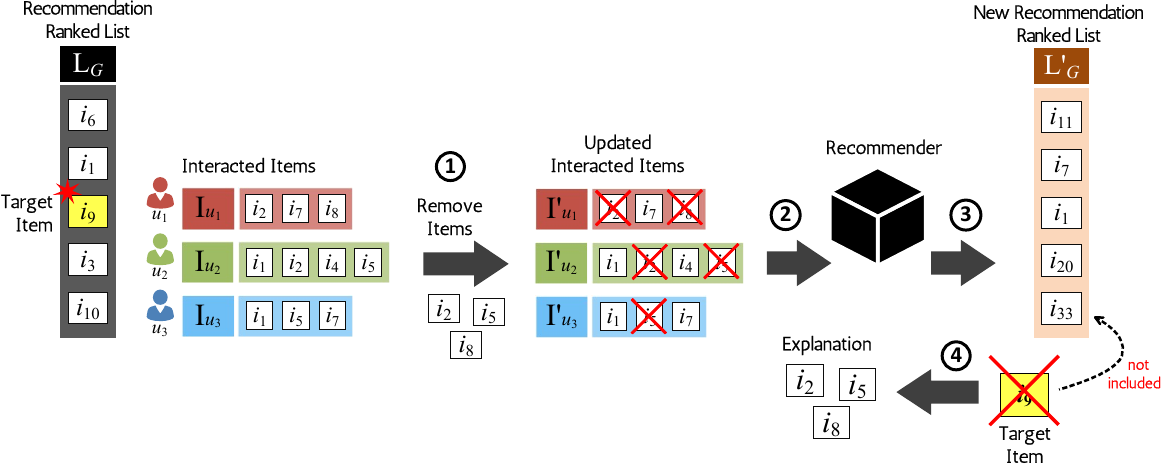}
     \caption{Explanation Process: Finding Counterfactual Explanations}
     \label{fig:approach}
 \end{figure*}

A \textit{counterfactual explanation for a group recommendation} identifies which past interactions of the group members are responsible for the presence of a particular item (the \textit{target item}) in the group's recommendation list.
To find such an explanation, we consider a set of items with which the group members have previously interacted and simulate their removal from the group's interaction history. We then re-run the group recommender and examine whether the target item still appears in the resulting recommendation list. If the target item no longer appears, the removed items constitute a valid counterfactual explanation, as their removal is sufficient to change the recommendation outcome. The overall explanation process is outlined in Figure~\ref{fig:approach} (more details in Sec.~\ref{sec:model}).
The problem can be illustrated through the following question and its corresponding explanation.

 \begin{tcolorbox}
	[colback=gray!10,colframe=white,arc=0pt,outer arc=0pt,
left=4pt, right=4pt, top=6pt, bottom=6pt]
\sffamily\small
\begin{center}
\textbf{Question}

``\textit{Why does the target item $t$ appear in the recommendation list}?''

\vspace{6pt}

\textbf{Explanation}

``\textit{If you had not interacted with the items $X$, item $t$ would not have appeared in the recommendation list}.''
\end{center}
\end{tcolorbox}

One major advantage of our approach is that the explanations are formed from data items with which the group has previously interacted. This ensures that explanations remain grounded in the system’s internal knowledge and do not rely on external information, thereby \textit{preserving user trust}.
Moreover, our approach is \textit{model-agnostic and applicable to black-box recommender systems}, as it does not require access to their internal workings, but only to their outputs.

However, finding explanations is computationally challenging, particularly for groups. Since a group consists of multiple users, each with potentially numerous interactions, the set of candidate items to be examined is large.
Furthermore, identifying an explanation requires examining all the different combinations of these items, whose number grows exponentially with respect to the number of items.
As a result, exhaustive search becomes prohibitively costly, especially since evaluating each candidate subset requires invoking the group recommender.

To this end, we propose a framework comprising five efficient algorithms   that employ different techniques, including search-space pruning, greedy selection, early stopping, backward pruning, and incremental refinement.
These techniques aim to reduce the number of candidate explanations examined and the required recommender calls, while considering different trade-offs between computational efficiency and explanation quality.
Moreover, we introduce a set of metrics that capture different dimensions of item relevance, including the extent of the group’s interactions with an item, its system-wide popularity, its contribution to removing the target item from the recommendation list, and group members’ preferences for the target item.

Specifically, we design five algorithms that aim to maximize explanation utility while minimizing computational cost.
First, the \textit{Pareto-filtering} algorithm exploits Pareto dominance to prune candidate items and reduce the search space, requiring a limited number of recommender calls.
The \textit{GreedyGrow} algorithm ranks candidate items based on their scores and incrementally adds them until a valid counterfactual explanation is found.
The \textit{ExpRebuild} and \textit{Grow\&Prune} algorithms extend \textit{GreedyGrow} by introducing additional refinement steps.
In particular, the \textit{ExpRebuild} algorithm  orders items according to their ability to explain the target item and then incrementally reconstructs the counterfactual.
By contrast, \textit{Grow\&Prune} starts from an initial counterfactual and iteratively removes unnecessary items to produce a more concise explanation. Notably, explanation size is a crucial factor in explanation clarity \cite{abs-1802-00682,KliegrBF21} (Sec.~\ref{sec:cf_metrics}).
Finally, \textit{FixedWindow} applies a sliding window over the ranked candidate items to identify a valid counterfactual, and then performs an exhaustive search within the restricted window to find a more compact explanation.

Together, these algorithms explore different trade-offs among computational efficiency, explanation size, utility, and fairness, which we evaluate in our experiments.

\stitle{Contribution.}
The main contributions are the following:

\begin{itemize}
   \item \textbf{Problem Definition.}
We introduce and formally define the problem of finding counterfactual explanations in group recommender systems.

\item \textbf{Item and Explanation Metrics.}
We introduce a comprehensive set of item-level metrics that assess recognition, likeness, expressivity, and influence, both within and beyond the group.
Additionally, we define a set of explanation-level metrics that evaluate desirable properties, such as clarity, interpretability, cost, and fairness.

\item \textbf{Algorithms.}
We design five algorithms for efficiently discovering group counterfactual explanations, with different trade-offs regarding computational efficiency,  cost of recommender calls, and explanation quality. We also provide a thorough complexity analysis of each algorithm in terms of computational time, the number of recommender calls, and space requirements.

\item \textbf{Experimental Evaluation.}
We evaluate the proposed algorithms on two real-world datasets (MovieLens and Amazon). Our results highlight the trade-offs among the algorithms in terms of explanation quality, explanation size, computational cost, and fairness.
\end{itemize}

\stitle{Previous Work.}
A preliminary version of this work appeared at the International Workshop on
Design, Optimization, Languages and Analytical Processing of Big
Data (DOLAP) 2025~\cite{DBLP:conf/dolap/StratigiBS25}.
The present paper substantially extends that study in three directions.
$(1)$~\textit{We considerably extend the algorithmic framework} by refining the original Sliding Window method into FixedWindow, and introducing three additional algorithms (GreedyGrow, Grow\&Prune, and \mbox{ExpRebuild}), and developing a Pareto-based filtering mechanism to reduce the search space.
$(2)$~\textit{We introduce a richer set of item-level and explanation-level metrics}, including new measures of interpretability, utility, and fairness.
$(3)$~\textit{We substantially expand the experimental evaluation} by replacing MovieLens 100K with MovieLens 10M and introducing an additional large-scale dataset, Amazon.
Moreover, we provide a more comprehensive experimental analysis, examining several new factors and parameters,  such as computational cost, explanation size, interpretability, utility, and fairness.

\stitle{Outline.}
The remainder of this paper is organized as follows.
Section~\ref{sec:model} presents the basic concepts of the counterfactual explanation model.
Then, Section~\ref{sec:item_metrics} introduces the item-level metrics.
Section~\ref{sec:utility} presents explanation-level metrics and defines the problem.
Next, Section~\ref{sec:algorithmic_framework} presents and analyzes the proposed algorithms. Section~\ref{sec:exp} outlines the experimental setup and discusses the evaluation results.
Section~\ref{sec:relatedWork} reviews related work.
Finally, Section~\ref{sec:conc} concludes the paper.

\begin{table}[t]
\centering
\footnotesize
\setlength{\tabcolsep}{4pt}
\setlength\extrarowheight{4pt}
\caption{Summary of Common Notations}
\label{tab:notations}
\begin{tabularx}{\linewidth}{@{}lX@{}}
\toprule
\textbf{Notation} & \textbf{Description} \\
\dlineB

\tabsection{Users and Items}

$U$, $u$ & Set of all users, a user \\
\rowcolor{gray!15}
$G$ & A group of users \\
$I$, $i$ & Set of all items, an item \\
\rowcolor{gray!15}
$t$ & Target item, i.e., the item to be explained \\
$I_u$ & Items with which user $u$ has interacted \\
\rowcolor{gray!15}
$I_G$ & Items with which any member of $G$ has interacted \\
$\rho_{i,u}$ & Rating assigned by user $u$ to item $i$ \\

\tabsection{Recommendations}

$\pi(I)$ & Group recommendation system given item set $I$ \\
\rowcolor{gray!15}
$L_I$ & Group ranked recommendation list produced from $I$ \\
$\fnc{rank}(i, L_I)$ & Rank position of item $i$ in list $L_I$ \\

\tabsection{Explanations and Metrics}

$E$ & Counterfactual explanation \\
\rowcolor{gray!15}
$\fnc{rc}(i, S)$ & Average recognition of item $i$ among users $S$ \\
$\fnc{rt}(i, S)$ & Average rating of item $i$ by users $S$ \\
\rowcolor{gray!15}
$\fnc{expwr}(P, t, I)$ & Explanatory power of item subset $P$ with respect to target item $t$ and item set $I$ \\
$\fnc{infl}(i, t, I, G)$ & Influence of item $i$ on target item $t$ for group $G$ \\
\rowcolor{gray!15}
$i.\textit{score}$ & Total score of item $i$ \\
$\fnc{fair}(G, E)$ & Fairness of explanation $E$ with respect to group~$G$ \\
\rowcolor{gray!15}
$\fnc{min}(E)$ & Minimality of explanation $E$ \\
$\fnc{interpr}(E)$ & Interpretability of explanation $E$ \\
\rowcolor{gray!15}
$\fnc{cost}(E)$ & Computational cost of explanation $E$ \\
$\Omega(E)$ & Utility of explanation $E$ \\
\bline
\end{tabularx}
\end{table}

\section{Group Counterfactual Explanations Model}
\label{sec:model}

To enhance the transparency and interpretability of group recommendation systems, we introduce a model for generating counterfactual explanations tailored to groups of users. These explanations provide actionable insights by identifying which items would lead to different recommendation outcomes if removed from the group’s interaction history. In this section, we present the foundational concepts of our model, formalize the definition of group counterfactual explanations, and introduce relevant item-level metrics to support the explanation generation process. Table~\ref{tab:notations} summarizes the notation used throughout the paper.

\subsection{Basic Concepts}
\label{sec:model_concept}

We begin by defining the core entities involved in our model, namely users, groups, their interactions with items, and how these interactions shape group recommendations.

\stitle{User, Group \& Interacted Items.} Let $U$ denote the set of all \textit{users} in the system.
For each user $u \in U$, let $I_u$ represent the set of \textit{items that the user has interacted with}, such as items they have ordered, rated, clicked, or liked.
Let $G \subseteq U$ be a group of users, and let $I_G$ denote the set of \textit{items that any group member has interacted with}, formally defined as $I_G = \bigcup_{\forall u \in G} I_u$.
Furthermore, $\rho_{i,u} \in [0,1]$ represents the \textit{rating assigned by user} $u$ to \textit{item} $i$.
If a rating is not present, it is assumed to be zero. It is important to note that our approach does not require the presence of ratings for all users or items.

\stitle{Recommendations.} Let $\pi(I)$ denote a \textit{group recommendation system}, which, given a set of interacted items $I$, produces an ordered \textit{recommendation list} $L_I$, i.e.,~${\pi(I) = L_I}$.
For a given recommendation list $L_I$, let $\fnc{rank}(i, L_I) \in [1, |L_I|]$ denote the \textit{position (rank) of item} $i$ in $L_I$.
Generating a recommendation list $L_I$ is associated with a \textit{recommender call cost}~$\phi$, such as system latency or API usage cost.

In this work, we treat the group recommender system as a \textit{black box}, enabling general applicability of our approach regardless of the internal logic or type of recommendation algorithm employed.

\subsection{Group Counterfactual Explanations}
Having defined the basic entities and the recommendation mechanism, we now formalize the concept of counterfactual explanations in group settings.
Inspired by prior work on counterfactuals in individual recommendations \cite{10.1145/3450613.3456846}, we extend the concept to group scenarios, where explanations must account for collective interactions rather than those of a single user.

The essence of a counterfactual explanation lies in identifying the set of previously interacted items that, if removed, would prevent the appearance of a specific item (the \textit{target item}) in the recommendation list.






\stitle{Group Counterfactual Explanations.}
Formally, a \textit{group counterfactual explanation} is defined as follows.
Given a group $G \subseteq U$ and a target item $t \in L_{I_G}$, a \textit{group counterfactual explanation} $E \subset I_G$ is a set of interacted items such that if the group had not interacted with these items, the item $t$ would not have been recommended\footnote {For the remainder of this paper, \textit{Counterfactual explanation} will be referred to interchangeably as \textit{Counterfactual} or \textit{Explanation}.}.

Considering the above, the following holds.
Let \linebreak
${\pi(I_G) = L_{I_G}}$ with $t \in L_{I_G}$.
Then, a set of items $E \subset I_G$ is a \textit{counterfactual explanation} if
$\pi(I_G \backslash E) = L_{I'_G}$ and $t \notin L_{I'_G}$.

\stitle{Factual and Counterfactual Scenario.}
In this setting, the \textit{factual scenario} is defined by the current interaction set $I_G$, where $\pi(I_G) = L_{I_G}$ and $t \in L_{I_G}$.
The \textit{counterfactual scenario} involves a modified interaction set $I'_G = I_G \backslash E$, such that $\pi(I'_G) = L_{I'_G}$ and $t \notin L_{I'_G}$.
The difference ${E = I_G \backslash I'_G}$ thus constitutes the counterfactual explanation.
This approach inherently assumes that the only \textit{feasible action} is the \textit{removal of items} from the group’s interaction history.

\stitle{Explanation Process Example.}
Figure~\ref{fig:approach} illustrates the overall explanation process.
Suppose  the recommendation list  is $L_G$ and the \textit{target item} is $i_9$, i.e., the item for which we aim to generate an explanation.
The group consists of \textit{three users}: $u_1$, $u_2$, and $u_3$, each associated with their respective \textit{interaction lists}, denoted as $I_{u_1}$, $I_{u_2}$, and $I_{u_3}$.
\circlednum{1}~Initially, we remove a selected set of items from the users' interaction lists, i.e., $i_2$, $i_5$, and $i_8$. This set represents a candidate counterfactual explanation.
\circlednum{2}~Next, we input the modified interaction lists $I'_{u_1}$, $I'_{u_2}$, and $I'_{u_3}$ into the recommendation system.
\circlednum{3}~The system returns a new group recommendation list $L'_G$.
\circlednum{4}~Since the target item $i_9$ no longer appears in $L'_G$, the removed items $i_2$, $i_5$, and $i_8$ constitute a counterfactual explanation.

\section{Item Metrics}
\label{sec:item_metrics}

To support the generation and evaluation of counterfactual explanations, we define a set of item-level metrics that quantify various aspects of item influence and group interaction patterns.

\stitle{Item Recognition.}
We begin by defining how widely the users know or recognize an item.
The \textit{item recognition} metric captures the fraction of users in a given set who have interacted with the item.

Let $i$ be an item and $S \subseteq U$ a set of users. Then, item recognition $\fnc{rc}(i, S) \in [0,1]$ is defined as:
%
\begin{equation}
    \fnc{rc}(i, S)=\frac{
    \underset{\substack{\forall u \in S}}{\sum} \zeta(u,i)}{|S|}
    \label{eq:item_rec}
\end{equation}

\noindent
where $\zeta(u,i) = 1$ if user $u$ has interacted with item $i$
\linebreak
(i.e., $i \in I_u$), and $0$ otherwise.

We evaluate this item recognition  in two contexts:
\linebreak
(1)~$\fnc{rc}(i, G)$: among the group members (\textit{group item recognition});  and
(2)~$\fnc{rc}(i, U \backslash G)$: among users outside the group \textit{(public item recognition)}.

\stitle{Item Rating.}
Next, we consider the average quality assessment of an item based on user feedback.
Given an item $i$ and a user set $S \subseteq U$, the \textit{item rating} is the average of ratings provided by $S$ users:
\begin{equation}
 \fnc{rt}(i, S) =  \frac{
      \underset{\forall u \in S}{\sum}\rho_{i,u} }{|S|}
    \label{eq:item_rat}
\end{equation}

Similar to recognition, item rating is computed in two contexts:
(1) $\fnc{rt}(i, G)$: among the  group members \textit{(group item rating)}; and
(2) $\fnc{rt}(i, U \backslash G)$: among all users excluding the group (\textit{public item rating}).

\stitle{Item Influence on the Target Item.}
The \textit{influence of an item on the target item} quantifies the extent to which a previously interacted item contributes to the recommendation score assigned to the target item $t$ by the recommender system, in a \textit{model-agnostic} setting.

Intuitively, interacted items that lead to a higher recommendation score for $t$, resulting in $t$ being ranked higher,  are considered to exert a stronger influence. These items are therefore more likely to appear in counterfactual explanations, as their removal can significantly alter the ranking of~$t$.

To estimate this influence in a model-agnostic way, we treat each user in the group individually and evaluate the effect of their interactions on the recommendation score of~$t$. Specifically, for each user $u \in G$, we call the group recommender using only their interaction history $I_u$ and retrieve the recommendation score for $t$.

Formally, let $G' = \{ u \in G \mid i \in I_u \}$ be the set of group members who have interacted with item $i$. Then, the \textit{influence} $\fnc{infl}(i,t,I, G) \in [0,1]$ of item $i$ on target item $t$ for group $G$ is computed as the average recommendation score of $t$ across $G'$  users:
\begin{equation}
\label{eq:item_infl}
    \fnc{infl}(i,t,I, G) = \sum_{\substack{
\forall u \in G' \\ I_u \subseteq I}} \!\! \fnc{recScore}(t, I_u) \Big/ |G'|
\end{equation}

\noindent
Here, $\fnc{recScore}(t, I_u) \in [0,1]$ denotes the recommendation score assigned to $t$ by the recommender system when invoked with only user $u$ interacted items $I_u$. This approach allows us to estimate item influence without requiring access to the internal workings of the recommender (i.e., in a \textit{model-agnostic} way).

\stitle{Item Set Explanatory Power.}
We introduce the \textit{explanatory power} metric, which quantifies the potential of an item set to contribute to a counterfactual explanation and, consequently, its ability to explain the presence of a target item. The metric is computed in a \textit{model-agnostic} manner, relying solely on observable changes in the recommendation output.

Let $I$ be the interacted items set, $L_I$ the resulting recommendation list with $|L_I| = m$, and $t$ the target item.
Let $P \subset I$ be a subset of items considered for removal, and let $\pi(I \backslash P)$ be the new recommendation list.

The \textit{explanatory power} of item set $P$, denoted \linebreak ${\fnc{expwr}(P, t, I) \in [0,1]}$, is defined as:

\begin{equation}
\fnc{expwr}(P,t,I) =
\begin{cases}
1, & \text{if } t \notin \pi(I\backslash P),\\
\frac{\fnc{rank}(t,\pi(I\backslash P))-1}{m},
& \text{otherwise}.
\end{cases}
\label{eq:item_power}
\end{equation}

Explanatory power quantifies the rank of $t$ when the item set $P$ is removed from the interacted item set $I$.
Higher values indicate that the items in  $P$ are more likely to be part of an explanation.

Particularly, if $t$ is eliminated from the list ($t \notin \pi(I \backslash P)$), then $P$ forms a counterfactual and its explanatory power is set to one.
On the other hand, if $P$ results in item $t$ being ranked top (i.e., top-1), explanatory power takes its lowest value of zero.

\stitle{Item Total Score.}
While individual metrics (e.g., recognition, rating, influence) offer valuable insights, combining them into a single score enables more effective ranking and selection of items for explanation. To this end, we define the \textit{item total score} as an aggregate of multiple item-based metrics.

Given an item $i$ and a collection of $k$ metrics
\linebreak
$(f_1, f_2, \dots, f_k)$, we compute the total score as:
\begin{equation}
    i.score = \sum_{j=1}^k f_j(i)
\end{equation}

\noindent
In this work, we use the following five metrics to calculate the total score:
\begin{enumerate*}[label=(\arabic*), itemjoin={{; }}, itemjoin*={{; and }}]
    \item $\fnc{rc}(i, G)$: group item recognition
    \item $\fnc{rc}(i, U \backslash G)$: public item recognition
    \item $\fnc{rt}(i, G)$: group item rating
    \item $\fnc{rt}(i, U \backslash G)$: public item rating
    \item $\fnc{infl}(i,t, I, G)$: item influence on the target item.
\end{enumerate*}

 Putting it all together, the final formula for the \textit{item total score} is:
\begin{equation}
\label{eq:totalscore}
\begin{aligned}
i.score ={}&
    \fnc{rc}(i, G)
    + \fnc{rc}(i, U\backslash G)
    + \fnc{rt}(i, G) \\
&\quad
    + \fnc{rt}(i, U\backslash G)
    + \fnc{infl}(i,t, I, G)
\end{aligned}
\end{equation}

This score serves as a unified measure of an item’s importance, considering both its relevance within the group and its potential impact on the target recommendation outcome. It plays a central role in guiding the selection of ``explanatory'' items in our counterfactual framework.

\section{Counterfactual Metrics \& Problem Definition}
\label{sec:utility}
In this section, we present a set of evaluation dimensions for assessing the quality of counterfactual explanations in group recommender systems.
Although prior research on counterfactual reasoning has largely centered on individual-level explanations, extending these approaches to group settings presents unique challenges and design considerations.
Beyond generating explanations that are minimal and interpretable, we must also consider the computational cost of discovering them, the overall utility they offer to the group, and the fairness of how explanatory changes are distributed among group members.

We begin by formalizing utility-oriented metrics that capture properties such as explanation minimality, interpretability, and generation cost, and introduce a unified utility function that balances these aspects. We then define a notion of counterfactual fairness tailored to groups, ensuring that no single user is disproportionately affected by the changes recommended in the explanation. Collectively, these metrics enable a comprehensive assessment of group-level counterfactual explanations, emphasizing clarity, utility, and fairness.

\vspace{6pt}
\subsection{Counterfactual Metrics}
\label{sec:cf_metrics}

In this section, we introduce a set of  metrics to assess the quality and characteristics of group-level counterfactual explanations in recommender systems. These metrics capture different desirable properties of explanations, including clarity, interpretability, cost-efficiency, and fairness.

\stitle{Counterfactual Explanation Minimality.}
We use the notion of counterfactual \textit{minimality} (also referred to as \textit{sparsity}) to evaluate the \textit{clarity} of an explanation; i.e., how easily it can be understood. Numerous studies in cognitive psychology and explainable AI have shown that explanations involving fewer changes are more understandable and cognitively efficient to process \cite{abs-1802-00682,KliegrBF21,Reiter87,Miller19,WangYAL19,2020_WSDM_GBRW,2017_Arxiv_WMR}.

In our setting, the minimality of a counterfactual explanation $E$ is quantified by the number of items it contains. A smaller $|E|$ indicates a clearer explanation. This quantity corresponds to the number of user-item interactions removed to achieve the counterfactual.

\vspace{10pt}
We formally define the \textit{counterfactual explanation minimality} as:

\begin{equation}
    \fnc{min}(E) = 1 - \frac{|E|}{|I_G|}
\end{equation}

\noindent where $|I_G|$ is the total number of interacted items by the group. The value is normalized in $[0,1]$ such that higher values indicate better minimality.

\stitle{Counterfactual  Explanation Interpretability.}
An explanation is more interpretable when the items it involves are familiar to users \cite{celar2023reasoning}. We measure interpretability through item recognition, i.e., how often an item appears in user interaction histories. This includes both (1) the recognition of an item within the group, and (2) its recognition by the broader user base.

To quantify this, we define the interpretability of a counterfactual explanation $E$ as the average recognition of its items:
\begin{equation}
\fnc{interpr}(E)   =
\frac{1}{2|E|} \Big(
\underset{\forall i \in E}{\sum}     \fnc{rc}(i, G)
    +
    \underset{\forall i \in E}{\sum}
    \fnc{rc}(i, U \backslash G)
    \Big)
\label{eq:recognition}
\end{equation}

\noindent Here, $\fnc{rc}(i, G)$ denotes the recognition of item $i$ in the group, and $\fnc{rc}(i, U \backslash G)$ represents its recognition among other users.

 \stitle{Counterfactual Explanation Cost.}
Producing a counterfactual explanation involves making multiple calls to the recommender system, each associated with a cost $\phi$. The total cost of an explanation $E$ is thus defined as the number of calls made to discover it:
\begin{equation}
    \fnc{cost}(E)=\sum_{i=1}^{n}\phi
\end{equation}

\noindent where $n$ is the number of recommender invocations required. For simplicity, we assume $\phi = 1$ in our evaluation, which allows us to count the number of calls directly.

\stitle{Counterfactual Explanation Utility.}
We define the \textit{utility} of a counterfactual explanation $E$ as a combination of its clarity and interpretability. The utility function is defined as:

\begin{equation}
\label{eq:expl_util}
\Omega(E)=\psi\big( \fnc{min}(E), \fnc{interpr}(E) \big)
\end{equation}

\noindent where $\psi$ is an aggregation function that balances the two metrics, for instance via a weighted sum or product.

\subsection{Problem Definition}
\label{sec:prob_def}

 We formally define the problem of finding \textit{Group Counterfactual Explanation} as follows.

\stitle{Group Counterfactual Explanation Problem.}
 Given a \textit{group} $G$;
 a \textit{group interacted items} $I_G$;
 a \textit{group recommended items list}  $L_{I_G}$;
 a \textit{target item} $t$;
   a \textit{recommender call cost} $\phi$;
 and a \textit{budget} $B$ in terms of recommender call cost;
our goal is to find a \textit{group counterfactual explanation} $E^*$,
 such that the \textit{explanation utility} $\Omega(E^*)$ is maximized and the
 \textit{counterfactual cost} $\fnc{cost}(E^*)$ does not exceed the budget $B$.

 \[
 	E^* =   \argmax \Omega(E) \quad \text{s.t.}  \quad  \fnc{cost}(E) \leq B
\]

\stitle{Computational Complexity.}
Solving this problem involves evaluating all subsets of $I_G$, i.e., the power set excluding the empty set. Therefore, the number of candidates is $2^{|I_G|} - 1$. Since each candidate requires a recommender system call with cost $\phi$, the computational complexity is $O(\phi \cdot 2^{|I_G|})$.

\subsection{Counterfactual Fairness}
\label{sec:cf_fairness}


The fairness notion used in this work is inspired by \textit{individual fairness}~\cite{DworkHPRZ12,BiegaGW18,KusnerLRS17}, which states that similar individuals should be treated similarly. We translate this to the group setting by requiring that the burden of a counterfactual explanation (i.e., the items removed) be fairly distributed among group members.

Formally, we say that an explanation $E$ is fair if each group member contributes a similar number of interacted items to the explanation. For any user $u \in G$, let $\zeta(u,i)$ be an indicator function equal to 1 if user $u$ has interacted with item $i$, and 0 otherwise.

To assess fairness, we compute the standard deviation of the number of items from each user's profile that are included in $E$:

\begin{align}
\fnc{fair}(G,E) &=
\frac{1}{
\xi+\sigma\left(
s_{u_1}, s_{u_2}, \dots, s_{u_{|G|}}
\right)
}
\label{eq:fair} \\[8pt]
s_u &=
\sum_{\forall i \in E} \zeta(u,i),
\qquad \text{where } u \in G. \notag
\end{align}

A lower standard deviation indicates more balanced participation and, therefore, greater fairness. By taking its inverse, we ensure that higher values of $\fnc{fair}(G,E)$ correspond to fairer explanations. A small constant $\xi>0$ is added to avoid division by zero. Perfect fairness is achieved when all users contribute equally, i.e., $\sigma = 0$.

Note that fairness is evaluated as an ex-post property of the generated explanations. The proposed algorithms do not explicitly optimize fairness; rather, we investigate how their different selection and refinement strategies affect the distribution of explanatory contributions among group members.

\section{Algorithmic Framework for Counterfactual Search}
\label{sec:algorithmic_framework}

The search space for group counterfactual explanations is exponential, making exhaustive search impractical. We therefore develop efficient algorithms that use several techniques  to  select and refine candidate explanations.
In the following sections we present five algorithms.
We first introduce \textit{ParetoFiltering} (Sec.~\ref{sec:pareto_algo}), which selects items based on multidimensional dominance and can serve as a preprocessing step, which   prunes the candidate item space early, requiring only a limited number of recommender calls.
\textit{FixedWindow}  (Sec.~\ref{sec:fixed_window})  scans a sorted candidate list using a sliding window to locate a counterfactual, within which an exhaustive local search is performed to extract a more compact explanation.
\textit{GreedyGrow} (Sec.~\ref{sec:greedy_grow}) follows a forward counterfactual construction strategy, progressively adding items until a counterfactual is found.
The following two methods refine this procedure.
\textit{Grow\&Prune} (Sec.~\ref{sec:grow_and_prune}), which starts from an initial counterfactual and iteratively eliminates items to reduce the explanation size;
and \mbox{\textit{ExpRebuild}} (Sec.~\ref{sec:influence_rebuild}), which ranks items by their ability to explain the target item and incrementally reconstructs the counterfactual.


\subsection{Search Space Reduction Using Pareto-based Filtering}
\label{sec:pareto_algo}
We first introduce \textbf{ParetoFiltering}, which uses Pareto dominance across item metrics to select candidate items. The key goal of  this method  is to improve efficiency by pruning dominated items before counterfactual search, thereby reducing the number of recommender calls required.
Hence, Pareto filtering can be considered as a preprocessing step to effectively reduce the search space.

Specifically, we leverage the item metrics introduced in Section~\ref{sec:item_metrics} to map each item into a multidimensional metric space, enabling dominance-based filtering to identify a ``high-quality'' counterfactual superset.

\stitle{Metric-based Representation.}
Let $d$ denote the number of item-level metrics. Each item $i_k \in I_G$ is represented as a $d$-dimensional vector $o_k = (o_k^1, o_k^2, \dots, o_k^d)$, where $o_k^j = f_j(i_k)$ is the value of metric $j$ for item $i_k$. The attribute space is defined over $\mathbb{R}^d$.

\stitle{Pareto Set.}
An object $o_1$ \textit{dominates} $o_2$, denoted $o_1 \succ o_2$, if:
(1)~$\forall w \in [1, d],\; o_1^w \geq o_2^w$, and
(2)~$\exists j \in [1, d],\; o_1^j > o_2^j$.

\noindent
The \textit{Pareto set}, or skyline \cite{KungLP75}, denoted $PS(O)$ for object set $O$, is defined as:
\[
PS(O) = \{ o_i \in O \mid \nexists o_k \in O : o_k \succ o_i \}.
\]
Items in $PS(O)$ are referred to as Pareto-optimal items and represent a non-dominated frontier within the metric space.

\stitle{Threshold-based Generalization.}
To flexibly control the cardinality of the result set, we generalize the Pareto set using a threshold vector $\tau = \langle \tau^1, \tau^2, \dots, \tau^d \rangle$ where $\tau^j \in \mathbb{R}$ adjusts strictness per metric.

\noindent
An object $o_1$ \textit{$\tau$-dominates} $o_2$, written $o_1 \succ_\tau o_2$, if:
\linebreak
(1)~$\forall w \in [1, d],\; o_1^w \geq o_2^w + \tau^w$, and
(2)~$\exists j \in [1, d],\; o_1^j > o_2^j$.
%
The $\tau$\textit{-Pareto set} is then defined as:
\[
PS_\tau(O) = \{ o_i \in O \mid \nexists o_k \in O : o_k \succ_\tau o_i \}.
\]

Setting $\tau = 0$ yields the standard Pareto set; $\tau < 0$ relaxes dominance criteria (reducing the size of $PS_\tau(O)$ set) while $\tau > 0$ tightens them (increasing the size of $PS_\tau(O)$ set). Efficient skyline algorithms like D\&C \cite{KungLP75}, SFS \cite{ChomickiGGL03}, LESS \cite{GodfreySG07}, and SaLSa \cite{BartoliniCP08} can be adapted to compute $PS_\tau(O)$.

\stitle{Pareto-based Filtering Technique Overview.}
To efficiently (i.e., using a small number of recommender calls) identify a counterfactual explanation set, we adopt an iterative filtering strategy based on $\tau$-Pareto dominance. Specifically, we remove from the group’s interaction items all items belonging to the $\tau$-Pareto set, which represent a collection of
“high-impact” items across the item-level metrics. The group recommender is then invoked to assess whether excluding these items causes the target item to disappear from the recommendation list. In such a case, the $\tau$-Pareto set is selected as the candidate counterfactual explanation.

In cases where the target item persists in the recommendation list, the examined $\tau$-Pareto set is not a counterfactual. To explore a broader set of items, we incrementally relax the dominance criteria by increasing the threshold vector $\tau$, and recompute the $\tau$-Pareto set.
This threshold-based relaxation enables a controlled expansion of the candidate item space. The process is repeated until a $\tau$-Pareto set corresponds to a counterfactual explanation.

\stitle{Algorithm Description.}
Algorithm~\ref{algo:pareto} takes as input an interacted item set $I$ and a target item $t$, and returns a counterfactual explanation $CS$.
Items are first mapped to $d$-dimensional metric vectors based on items metrics values (\textit{lines}~2--4).
The   main loop (\textit{lines}~7--14) iteratively relaxes the Pareto dominance criteria. At each iteration, a threshold vector $\tau$ is computed based on the standard deviation of each metric dimension (\textit{line}~9), and the $\tau$-Pareto set is computed (\textit{line}~10).
The items that correspond to the $\tau$-Pareto set objects are inserted into the item set $PI$ (\textit{line}~11).
The procedure \textsf{isCF} (Proc.~1) is then invoked to determine if $PI$ corresponds to a counterfactual explanation (\textit{line}~12).
Threshold relaxation continues until a counterfactual explanation is found.

\begin{algorithm2e}[t]

	\small
	\caption{ParetoFiltering  ( $I$, $t$ )}
	\label{algo:pareto}

	\KwIn{
		$I$: interacted items;  \:
		$t$: target item
	}
	\KwOut{
		$CS$: counterfactual explanation;
	}
	\KwPar{
		\mbox{$\pi$: group recommender;}  
		\mbox{$f_1, f_2, \dots, f_d$: item metrics functions}
	}

	\KwVar{
		$i$: item; \:
		\mbox{$O$:  \textit{d}-dimensional  objects;} \:
		\mbox{$it$:  iteration counter;} \:
		\mbox{$\tau$: attributes thresholds;} \:   
		\mbox{$PS$:   $\tau$-pareto set;} \:
		\mbox{$PI$:   items of  $\tau$-pareto set;}  \:
		\textit{cfFound}: boolean variable}

	\vspace{12pt}

	$O \gets \varnothing$

	\ForEach (\Comment*[f]{\mycomment{{transform items to \textit{d}-dimensional  objects}}}) {$i  \in I$}{

		$o_i \gets (f_1(i), f_2(i), \dots, f_d(i))$
		\Comment*[r]{\mycomment{{{compute the \textit{d} metrics for item $i$}}}}

		$O \gets O \cup o_i$
	}

	\vspace{4pt}

	$PI \gets \varnothing$

	$it \gets 0$

	\hidefornow{\note{\textbf{NIKOS TODO: find something to guarantee that PI==I can be achieved - avoid Infinite loop} }}

	\Do{
		\textnormal{(\textit{cfFound} $= \myfalse$\textnormal{)}}}{

		\lIf(\Comment*[f]{\mycomment{{ all items examined -- counterfactual not exist}}})
		{$PI = I$}{\textbf{break}}

		\vspace{4pt}

		$\tau \gets
			\Big\langle it \cdot \sigma(\underset{\forall o_i \in O}{\bigcup}o_i^1),
			it \cdot \sigma(\underset{\forall o_i \in O}{\bigcup}o_i^2), \dots,
			it \cdot \sigma(\underset{\forall o_i \in O}{\bigcup}o_i^d)
			\Big\rangle$
		\Comment*[l]{\mycomment{compute attributes' thresholds values as attributes std -- assign to $\tau$ the negative values}}

		\vspace{6pt}

		$PS  \gets$ \textnormal{find $\tau$-pareto set for objects $O$}

		$PI  \gets$ get the items of $I$ that correspond to  $\tau$-pareto set $PS$ objects
\Comment*[r]{\mycomment{{{items set to be checked as counterfactual}}}}

		\textit{cfFound} $\gets$ \textsf{isCF}$(I, PI, t)$
		\Comment*[r]{\mycomment{{{call recommender system}}}}

        $it \gets it+1$

	}

	\vspace{6pt}

	\eIf{cfFound $= \mytrue$}{
		$CS \gets PI$
	}(\Comment*[f]{\mycomment{{counterfactual not exist}}}){
		$CS \gets \varnothing$
	}

	\vspace{4pt}

	\Return $CS$

\end{algorithm2e}

\begin{procedure}[t]
	\small

	\SetAlgoProcName{\small Procedure 1: }{}
	\caption{isCF($I$, $S$, $t$)}

	\label{procedure:iscf}

	\KwIn{$I$: interacted items; \:
		$S$: interacted items to be checked as counterfactual; \:
		$t$: target item;

	}
	\KwOut{boolean value:(if $S$ is counterfactual)}

	\KwPar{
		\mbox{$\pi$: group recommender}}

	\KwVar{
		$L$: recommendation list generated considering $I \backslash S$ as interacted 
	}

	\vspace{8pt}

	$L \gets \pi(I \backslash S)$
	\Comment*[r]{\mycomment{{{call recommender system}}}}

	\vspace{4pt}

	\eIf{$t \notin L$}{
		\Return \mytrue
	}{
		\Return \myfalse
	}

\end{procedure}
\stitle{{Complexity: Time, Space, and Recommender Calls.}}

\vspace{-\bigskipamount}
\sititle{Time Complexity}:
In Algorithm~\ref{algo:pareto}, \textit{line}~3, considering the item metrics used in our implementation, the worst-case computational cost for computing the \textit{total item score} (Eq.~\ref{eq:totalscore}) is analyzed as follows.

The cost of computing the two item \textit{recognition metrics}, $\fnc{rc}(i, G)$ and $\fnc{rc}(i, U \backslash G)$ (Eq.~\ref{eq:item_rec}), is
 $O(max(|U|, |I|))$ per item (if hash-based indexes are used).
For simplicity, let us assume that $|U| > |I|$, so the complexity is $O(|U|)$.
The same holds for the \textit{rating metrics}, $\fnc{rt}(i, G)$ and $\fnc{rt}(i, U \backslash G)$ (Eq.~\ref{eq:item_rat}); both require $O(|U|)$ per item.
In order to compute the \textit{influence metric}, $\fnc{infl}(i, t, I, G)$ (Eq.~\ref{eq:item_infl}), for each user in $G$, we need to invoke the recommender.
Hence, the cost of computing influence metric is $O(|U|\phi)$, where $\phi$ is the recommender invocation cost.

Therefore, the total cost for computing all metrics for a single item is $O(|U| \phi)$, and for all items~$I$ is:
$O(|I||U| \phi)$.

In the \textit{second loop} (\textit{line}~7), the standard deviation computation in \textit{line}~9 costs $O(d|I|)$.
In \textit{line}~10 the $\tau$-pareto set computation  has a worst-case complexity of $O(|I| \log^{d-2} |I|)$ using partition-based algorithms~\cite{KungLP75}, or $O(d|I|^2)$ for practical algorithms such as SFS~\cite{ChomickiGGL03}, LESS~\cite{GodfreySG07}, or SaLSa~\cite{BartoliniCP08}.

The procedure \textsf{isCF} in \textit{line}~13 incurs a cost of $O(\phi)$.
The loop (\textit{line}~7) is executed up to a    small constant $\nu$ times, as the thresholds $\tau$ quickly converge to include all items in $I$. Therefore, the second loop (\textit{line}~7) has a worst-case complexity of:
$O\left( \nu \cdot (d|I| + |I| \log^{d-2} |I| + \phi) \right)$.

 Combining both loops, the overall \textit{worst-case time complexity} of the \textit{Pareto-filtering} algorithm is:
\[
O\left( |I||U| \phi  + \nu \cdot (|I| \log^{d-2} |I| + \phi) \right)
\]

\sititle{Space Complexity}:
The \textit{space complexity} of the algorithm (excluding input data) is $O(|I|)$, accounting for internal structures $O$, $PS$, $PI$, and $CS$.

\sititle{Recommender Calls}:
The \textit{number of recommender calls} in the worst case is $O(|I||U|+\nu)$.

\subsection{FixedWindow Search}
\label{sec:fixed_window}

The \textbf{FixedWindow} search technique serves as a local search strategy that incrementally examines contiguous subsets of   items.
The algorithm operates on a ranked item list and iteratively explores fixed-size windows, followed by an internal refinement phase to extract a small subset that corresponds to a counterfactual.

The interacted items are sorted by their total score, and a sliding window of fixed size is applied to the ordered list.
The window starts at the beginning of the list and moves one position at a time, covering consecutive items in each step. At every position, the algorithm simulates the removal of the window items from the group's interaction items and then re-invokes the group recommender system. If the target item is no longer recommended to the group, the current window corresponds to an explanation.
If the window traverses the entire list without yielding a counterfactual explanation, the window size is increased, and the list is scanned again from the beginning.

When the items within a window form a counterfactual explanation, we apply a \textit{refinement phase} to improve minimality (i.e., reduce the explanation size). Specifically, we examine all possible subsets (i.e., the powerset) of the window items in ascending order of cardinality. The first subset that corresponds to a counterfactual is selected as the final explanation, ensuring conciseness and interpretability.
Note that, since this refinement phase has exponential complexity, we keep the window size relatively small to balance explanation quality with computational efficiency.

\begin{algorithm2e}[t]
	\small
	\caption{FixedWindow($I$,  $t$, $w$)}
	\label{algo:fixedwindow}
	
	\KwIn{
        $I$:  interacted items;  \:
		$t$: target item; \:  
		$w$: initial window size
	}
	\KwOut{
		$CF$: counterfactual explanation
	}
	\KwPar{
		\mbox{$\pi$: group recommender;}  
		\mbox{$f_1, f_2, \dots, f_d$: item metric functions}
	}
	\KwVar{
		$H$: ranked item list; \:
		$W$: current window; \: 
		$S$: subset of window; \: \linebreak
		$w_{cur}$: current window size
	}

	\vspace{4mm}

    	$H \gets \varnothing$

	\ForEach(\Comment*[f]{\mycomment{{compute items scores \& initialize items list}}}) {$i \in I$}{

		$i.\textit{score} \gets \sum_{g=1}^d f_g(i)$
		\Comment*[r]{\mycomment{{{compute item score by aggregating item metrics}}}}

		insert $i$ into $H$
	}

	\vspace{4pt}

	sort $H$ in descending order based on item scores

	\vspace{4pt}


	$w_{cur} \gets \min(w, |H|)$ \Comment*[f]{\mycomment{{start with initial window size}}}

	\While{$w_{cur} \leq |H|$}{
		\For(\Comment*[f]{\mycomment{{sliding window search}}}){$i \leftarrow 1$ \KwTo $|H| - w_{cur} + 1$}{
			$W \gets H[i : i + w_{cur} - 1]$ \Comment*[r]{\mycomment{{window of $w_{cur}$ items}}}

			\If(\Comment*[f]{\mycomment{{check if window is a counterfactual}}}){\textnormal{\textsf{isCF}}$(I, W, t)$}{

				\ForEach(\Comment*[f]{\mycomment{{\textbf{refinement phase}: check subsets}}}){$S \in$ \textnormal{\textsf{PowerSet}}$(W)$ in order of increasing $|S|$}{
					\If{\textnormal{\textsf{isCF}}$(I, S, t)$}{
						$CF \gets S$ 
                        
                        \textbf{break}
					}
				}

				\lIf{$CF \neq \varnothing$}{\Return $CF$}
			}
		}
	
    \lIf{$w_{cur} = |H|$}{\Return $\varnothing$}

$w_{cur} \gets \min(w_{cur} + w, |H|)$ \Comment*[f]{\mycomment{{increase window size by initial $w$}}}

    }

	\vspace{2pt}
    
	\Return $\varnothing$
\end{algorithm2e}

\stitle{Algorithm Description.}
Algorithm~\ref{algo:fixedwindow} receives as input an interacted items set $I$, a target item $t$, and an initial window size $w$, and returns a counterfactual explanation $CF$.
Initially, the total item score is computed for each item in $I$ (\textit{lines}~2--4), and the items are sorted in descending order to form the ranked list $H$ (\textit{line}~5). Then, the current window size $w_{cur}$ is initialized to the predefined input value $w$ (\textit{line}~6).

For each window size, the algorithm slides a contiguous window $W$ over $H$ (\textit{line 9}) and checks whether the items $W$ correspond to a counterfactual explanation by invoking the \textsf{isCF} procedure (which invokes the recommender) (\textit{line}~10).

When a window contains a counterfactual explanation, the algorithm enters the refinement phase (\textit{lines}~11--14), where all non-empty subsets of $W$ are examined in increasing order of cardinality. The first subset $S$ that corresponds to a counterfactual is returned (\textit{line}~12--15).

If no counterfactual window is found, the window size is increased by $w$ and the search is repeated (\textit{line}~17). The algorithm terminates when either an explanation is identified  (\textit{line}~15) or all window sizes are examined (i.e., $w_{cur}=|H|$).

\stitle{{Complexity: Time, Space, and Recommender Calls.}}

\vspace{-\bigskipamount}
\sititle{Time Complexity}:
In Algorithm~\ref{algo:fixedwindow}, \textit{line}~3, each item $i \in I$ is assigned a score by aggregating the values of $d$ metric functions $f_1, f_2, \dots, f_d$. 
As shown in Section~\ref{sec:pareto_algo} the complexity to compute the metric functions for an item is $O(|U| \phi)$. Therefore, the cost of computing all scores is:
$O\left( |I| |U| \phi \right)$.

Sorting the ranked list $H$ (\textit{line}~5) employing a linearithmic sorting algorithm (e.g., mergesort) costs $O(|I| \log |I|)$.
In the main loop (\textit{lines}~7--17) 
%
the number of distinct window sizes considered in the worst case is  $\lceil |I|/w \rceil$. For each window size $w_{cur}$, the algorithm checks $O(|I|-w_{cur}+1)$ windows.
Hence, the total number of windows examined is
$\sum_{k=1}^{\left\lceil \frac{|I|}{w} \right\rceil}
\left( |I| - \min(kw,|I|) + 1 \right)$, which corresponds to:
$O\left( \frac{|I|^2}{w} \right)$.

For each window, the algorithm invokes the \textsf{isCF} procedure once (\textit{line}~10), incurring a cost of $O(\phi)$.  
%
In \textit{line} 11, the worst case, all 2$^{w_{cur}} - 1$ subsets are evaluated, each requiring a call to \textsf{isCF} with cost $O(\phi + m)$ (\textit{line}~12). Since $w_{cur}$ can grow up to $|I|$, the refinement phase may require up to $2^{|I|}$ checks in the worst case.  
Thus, the  computational cost of the main loop (\textit{lines}~7--17) is:
\[
O\left( \frac{|I|^2}{w} \phi  + 2^{|I|} \phi \right)
\]

 Combining all parts, the overall \textit{worst-case time complexity} of the \textit{FixedWindow} algorithm is:
\[
O\left( |I| |U| \phi + |I| \log |I| + \frac{|I|^2}{w}  \phi + 2^{|I|} \phi  \right)
\]

\sititle{Space Complexity}:
The \textit{space complexity} is $O(|I|)$, accounting for the sorted list $H$ and the temporary subset storage during refinement.

\sititle{Recommender Calls}:
The \textit{number of recommender calls} in the worst case is:
\[
O\left( |I||U|+\frac{|I|^2}{w} + 2^{|I|} \right)
\]

Note that, in practice, the exponential term is rarely reached due to computational cut-offs, and $w$ is typically chosen as a small constant, which keeps the quadratic term as the dominant factor.

\subsection{GreedyGrow Search}
\label{sec:greedy_grow}

The \textbf{GreedyGrow} search is a  method that progressively expands the search window until a counterfactual explanation is found. Unlike the fixed-size sliding window approach, this technique selects the top-ranked items in the list and adds one item at a time in sequential order.
At each iteration, the set of top-ranked items is checked to see if it corresponds to a counterfactual explanation.

This method is particularly effective at rapidly identifying explanations with strong influence; however, it has a notable limitation: the resulting explanations are not guaranteed to be minimal. Specifically, although the approach identifies a subset that successfully alters the recommendation outcome, it may include extraneous items that are not necessary for forming a counterfactual.
Consequently, the final explanation may be ``unnecessarily'' large, which, as discussed in Section~\ref{sec:utility}, can reduce interpretability and increase cognitive load for end users.

\stitle{Algorithm Description.}
Algorithm~\ref{algo:GreedyGrow} implements the \linebreak \mbox{\textit{GreedyGrow}} search strategy, takes as input the interacted item set $I$ and a target item $t$, and returns the explanation $CF$.
The algorithm first computes  the item total score for each interacted item and stores the items in a ranked list $H$ (\textit{lines}~2--5). It then initializes an empty candidate set $S$ (\textit{line}~6).

During each iteration of the main loop (\textit{lines}~7--10), the next highest-ranked item from $H$ is added to $S$, and the algorithm checks whether $S$ constitutes a counterfactual explanation by invoking \textsf{isCF} (Procedure 1). The process terminates as soon as the target item is removed from the recommendation list.

If a counterfactual explanation is found, the current set $S$ is returned as the final explanation (\textit{line}~12). Otherwise, if all items are exhausted without success, the algorithm returns the empty set (\textit{line}~14).

\begin{algorithm2e}[t]
	\small

	\caption{GreedyGrow   ($I$, $t$) }

	\label{algo:GreedyGrow}
	\KwIn{
		$I$:  interacted items;  \:
		$t$: target item;
	}
	\KwOut{
		$CF$: counterfactual explanation;

	}
	\KwPar{
		\mbox{$\pi$: group recommender;}  \:
		\mbox{$f_1, f_2, \dots, f_d$: item metrics functions;}
	}
	\KwVar{
		$H$: items list; \:
		$i$: item; \:
		$S$: selected items set; \: 
		\textit{cfFound}: boolean variable;
	}

	\vspace{4mm}

	$H \gets \varnothing$

	\ForEach(\Comment*[f]{\mycomment{{compute items scores \& initialize items list}}}) {$i \in I$}{

		$i.\textit{score} \gets \sum_{g=1}^d f_g(i)$
		\Comment*[r]{\mycomment{{{compute item score by aggregating item metrics}}}}

		insert $i$ into $H$
	}

	\vspace{4pt}

	sort $H$ in descending order based on item scores

	\vspace{4pt}

	$S \gets \varnothing$

	\For(\Comment*[f]{\mycomment{{select top items}}})
	{$j \leftarrow  1$ \KwTo $|H|$} {

		$S \gets S \cup H{[j]}$
\Comment*[r]{\mycomment{{{items set to be checked as counterfactual}}}}

		\textit{cfFound} $\gets$ \textsf{isCF}$(I, S, t)$
		\Comment*[r]{\mycomment{{{call recommender system}}}}

		\vspace{2pt}
		\lIf{cfFound $= \mytrue$}{ \textbf{break}}
	}

	\vspace{4pt}

	\eIf{cfFound $= \mytrue$}{
		$CF \gets S$
	}(\Comment*[f]{\mycomment{{no counterfactual exists}}}){
		$CF \gets \varnothing$
	}

	\vspace{4pt}

	\Return $CF$

\end{algorithm2e}

\stitle{{Complexity: Time, Space, and Recommender Calls.}}

\vspace{-\bigskipamount}
\sititle{Time Complexity}:
In  \textit{line}~3, each item $i \in I$ is assigned a score by aggregating the values of $d$ metric functions $f_1, f_2, \dots, f_d$.
As shown in Section~\ref{sec:pareto_algo}, the complexity to compute the metric functions for an item is $O(|U|  \phi)$. Thus, computing all item scores has a total cost:
$O\left( |I| |U| \phi \right)$.

Sorting the scored list $H$ (\textit{line}~6) takes $O(|I| \log |I|)$.
Then, in the main loop (\textit{lines}~7--10), items from $H$ are incrementally added to the set $S$. After each addition, the \textsf{isCF} function is called (\textit{line}~9) to check whether $S$ constitutes a valid counterfactual explanation. In the worst case, this process continues until all $|I|$ items have been added.
Each call to \textsf{isCF} has cost $O(\phi)$.
Thus, in the worst case, the total cost of the calls to \textsf{isCF} is:
$O\left( |I|\phi \right)$.

 From the above analysis, the overall \textit{worst-case time complexity} of the \textit{GreedyGrow} algorithm is:
\[
O\left( |I||U|\phi + |I| \log |I|  \right).
\]

\sititle{Space Complexity}:
The \textit{space complexity} is $O(|I|)$, due to the storage of the list $H$ and the growing set $S$.

\sititle{Recommender Calls}:
The \textit{number of recommender calls} in the worst case is $O(|I||U|+|I|)=O(|I||U|)$.

\subsection{Grow\&Prune Search}
\label{sec:grow_and_prune}

To reduce the size of explanations produced by \linebreak \mbox{GreedyGrow}, we introduce the \textbf{Grow\&Prune} search method, which first identifies an initial explanation and then refines it by improving minimality through backward elimination.

The \textit{Grow\&Prune} approach initially applies the \linebreak \mbox{\textit{GreedyGrow}} technique to generate an initial counterfactual explanation. Once this initial explanation is identified, we enter a \textit{reduction phase that evaluates the necessity of each item within the explanation}.
Each item in the initial counterfactual explanation is examined individually by temporarily excluding it from the set. For each exclusion, the group recommender is evaluated on the modified explanation to verify whether it continues to correspond to a counterfactual. If the exclusion results in a non-counterfactual explanation, the item is classified as essential and reinserted; otherwise, the exclusion is preserved, yielding a more concise explanation.

\stitle{Algorithm Description.}
Algorithm~\ref{algo:growprune} implements the
\linebreak
\textit{Grow\&Prune} strategy.
The algorithm first executes a grow phase (\mbox{\textit{lines}~1--10}) identical to the \textit{GreedyGrow} algorithm, incrementally adding items until a counterfactual explanation is identified. If no such explanation is found, the algorithm terminates and returns the empty set (\textit{line} 20).

Once a counterfactual explanation is obtained, the items in the candidate set $S$ are sorted in ascending order of score (\textit{line}~12), and the algorithm enters the prune phase. In this phase, each item $i \in S$ is temporarily removed from the explanation (\textit{line}~15), and the algorithm checks whether the reduced set still constitutes a valid counterfactual by invoking \textsf{isCF} (Proc.~1) (\textit{line}~16).

If the counterfactual condition is preserved, the removal is retained (\textit{line}~17); otherwise, the item is restored. The single-pass procedure continues until all items in the initial explanation have been examined once. The resulting set is returned as the final counterfactual explanation $CF$ (\textit{line}~18).

\begin{algorithm2e}[t]
	\small

	\caption{Grow\&Prune   ($I$, $t$) }

	\label{algo:growprune}
	\KwIn{
		$I$:  interacted items;  \:
		$t$: target item
	}
	\KwOut{
		$CF$: counterfactual explanation
	}
	\KwPar{
		\mbox{$\pi$: group recommender;} \:
		\mbox{$f_1, f_2, \dots, f_d$: item metrics functions}
	}
	\KwVar{
		$H$: items list; \:
		$i$: item; \:
		$S$: selected items set; \:
		\textit{cfFound}: boolean variable
	}

	\vspace{4mm}

	$H \gets \varnothing$

	\ForEach(\Comment*[f]{\mycomment{{compute items scores \& initialize items list}}}) {$i \in I$}{

		$i.\textit{score} \gets \sum_{g=1}^d f_g(i)$
		\Comment*[r]{\mycomment{{{compute item score by aggregating item metrics}}}}

		insert $i$ into $H$
	}

	sort $H$ in descending order based on items scores

	$S \gets \varnothing$

	\For(\Comment*[f]{\mycomment{{\textbf{grow phase}}}}){$j \leftarrow 1$ \KwTo $|H|$} {

		$S \gets S \cup H{[j]}$

		\textit{cfFound} $\gets$ \textsf{isCF}$(I, S, t)$

		\lIf{cfFound $= \mytrue$}{ \textbf{break}}
	}

	\eIf{cfFound $= \mytrue$}{
		
		sort $S$ in ascending order based on item scores

		$CF \gets S$

		\ForEach(\Comment*[f]{\mycomment{{\textbf{prune phase}}}}){$i \in S$}{
			
			$CF' \gets CF \setminus \{i\}$

			\If(\Comment*[f]{\mycomment{{check if $CF'$ still a counterfactual}}}){\textnormal{\textsf{isCF}}$(I, CF', t)$}{
				$CF \gets CF'$
			}
		}

		\Return $CF$
        	}(\Comment*[f]{\mycomment{{counterfactual not found}}}){
		\Return $\varnothing$
	}
\end{algorithm2e}

\stitle{{Complexity: Time, Space, and Recommender Calls.}}

\vspace{-\bigskipamount}
\sititle{Time Complexity}:
In \textit{line}~3, each item $i \in I$ is assigned a score via the aggregation of $d$ metric functions $f_1, \dots, f_d$.
As shown in Section~\ref{sec:pareto_algo}, the complexity to compute the metric functions for an item is $O(|U|  \phi)$. Thus, computing all item scores has a total cost:
$O\left( |I| |U| \phi \right)$.

Sorting the scored list $H$ (\textit{line}~6) takes $O(|I| \log |I|)$.
Then, during the \textit{grow phase} (\textit{lines}~7--10), the algorithm incrementally builds a candidate set $S$ by scanning items in $H$ in descending order. After each addition, the \textsf{isCF} function is invoked. In the worst case, this loop iterates $|I|$ times, leading to:
$O\left( |I| \cdot \phi  \right)$.

In \textit{prune phase} (\textit{lines}~11--20), the items in $S$ (of size at most $|I|$) are sorted in descending score order, costing at most $O(|I| \log |I|)$.
Then, for each item $i \in S$, the algorithm checks whether removing $i$ still yields a valid counterfactual by calling \textsf{isCF}. In the worst case, this requires $O(|I|)$ additional calls, resulting in:
$O\left( |I| \log |I|+|I| \phi \right)$.

 Combining all parts, the overall \textit{worst-case time complexity} of the  \textit{Grow\&Prune} algorithm is:
\[
\begin{aligned}
&O\left( |I| |U|  \phi + |I| \log |I| + |I| \phi + |I| \log |I|+|I| \phi \right) \\
&= O\left(|I| |U|  \phi + |I| \log |I| \right)
\end{aligned}
\]

\sititle{Space Complexity}:
The \textit{space complexity} is $O(|I|)$, due to the storage of lists $H$, $S$, and temporary counterfactual sets.

\sititle{Recommender Calls}:
The \textit{total number of recommender calls} in the worst case is $O(|I| |U|)$.

\subsection{ExpRebuild Search}
\label{sec:influence_rebuild}

The \textbf{ExpRebuild} search approach extends \textit{GreedyGrow} by incorporating \textit{items explanatory power} (Eq.~\ref{eq:item_power}) to guide item selection.
Specifically, exploiting explanatory power, it prioritizes items whose removal is expected to most strongly degrade the rank of the target item, or eliminate it entirely from the recommendation list.

Initially the method applies the \textit{GreedyGrow} procedure to efficiently find an initial counterfactual explanation.
During this process, the algorithm computes the explanatory power (Equation \ref{eq:item_power})
of each item.

After an initial explanation is identified, the items are sorted in descending order of explanatory power, placing the most impactful items first. The algorithm then constructs a new explanation incrementally: starting from an empty set, items are added one by one in order of decreasing explanatory power. After each addition, the recommender is invoked to check whether the growing set constitutes a counterfactual explanation.

This approach offers a compromise between minimality and interpretability, aiming to produce more compact explanations  grounded in a transparent attribution of item effects.

\begin{algorithm2e}[t]
	\small

	\caption{ExpRebuild   ($I$, $t$) }

	\label{algo:infuencerebuild}
	\KwIn{
		$I$:  interacted items;  \:
		$t$: target item;   \:
	}
	\KwOut{
		$CF$: counterfactual explanation;
	}
	\KwPar{
		\mbox{$\pi$: group recommender;} \:
		\mbox{$f_1, f_2, \dots, f_d$: item metrics functions}
	}
	\KwVar{
		$H$: items list; \:
		$i$: item; \:
		$S$: selected items set; \:
		\textit{inflScore}: influence score ; \:
		\textit{cfFound}: boolean variable; \:
		$P$: selected items from $S$ \:
	}

	\vspace{4mm}


	$H \gets \varnothing$

	\ForEach(\Comment*[f]{\mycomment{{compute items scores \& initialize items list}}}) {$i \in I$}{

		$i.\textit{score} \gets \sum_{g=1}^d f_g(i)$
		\Comment*[r]{\mycomment{{{compute item score by aggregating item metrics}}}}

		insert $i$ into $H$
	}

	\vspace{4pt}

	sort $H$ in descending order based on items scores

	\vspace{4pt}

	$S \gets \varnothing$


	\For(\Comment*[f]{\mycomment{{select top items}}})
	{$j \leftarrow  1$ \KwTo $|H|$} {


		$S \gets S \cup H{[j]}$
\Comment*[r]{\mycomment{{{items set to be checked as counterfactual}}}}

		\textit{cfFound},  \textit{inflScore}  $\gets$
		\textsf{checkForCFandComputeExplPwr}$(I, t, S)$
		\Comment*[r]{\mycomment{{{check if $S$ is counterfactual and compute $H[j]$ influence score}}}}

		$H{[j]}$.\textit{pow} $\gets$ \textit{inflScore}/|S|
        		  \Comment*[r]{\mycomment{{{set explanatory power score to item~$H{[j]}$ }}}}

		\vspace{2pt}
		\lIf{cfFound $= \mytrue$}{ \textbf{break}}
	}

	\vspace{4pt}

	\eIf(\Comment*[f]{\mycomment{{rebuild based on explanatory power}}}){cfFound $= \mytrue$}{

 sort $S$ in descending order based on the explanatory power score

		$P \gets \varnothing$

		\For
		{$j \leftarrow  1$ \KwTo $|S|$} {

			$P \gets P \cup S{[j]}$
\Comment*[r]{\mycomment{{{items set to be checked as counterfactual}}}}


				\textit{cfFound} $\gets$ \textsf{isCF}$(I, P, t)$
				

			\vspace{2pt}
			\lIf    {cfFound $= \mytrue$}{\Return $P$}
		}

		\Return S;

	}(\Comment*[f]{\mycomment{{counterfactual not exist}}}){
			\Return $\varnothing$
		}

	\vspace{4pt}


\end{algorithm2e}

\begin{procedure}[]
	\small

	\SetAlgoProcName{\small Procedure 2: }{}
	\caption{checkForCFandComputeExplPwr($I$, $t$,  $S$)}

	\label{procedure:iscf-infl}

	\KwIn{
    $I$: interacted items; \:
    $t$: target item; \:
	$S$: interacted items to be checked as counterfactual; \:
 		
	}

	\KwOut{
		$isCF$: boolean value (if $S$ is a counterfactual); \: 
		\textit{explPower}: explanatory power score of item~set~$S$
	}

	\KwPar{
		\mbox{$\pi$: group recommender}}

        \KwVar{
		$L$: recommendation list generated considering $I \backslash S$ as interacted}

	\vspace{8pt}


	$L \gets \pi(I \backslash S )$
    	\Comment*[r]{\mycomment{{{call recommender system}}}}


\eIf{$t \notin L$}{
    $isCF \gets \mytrue$

    $\textit{explPower} \gets 1$
}{
    $isCF \gets \myfalse$

    $\textit{explPower} \gets
    \frac{\fnc{rank}(t,L)-1}{|L|}$
    \Comment*[r]{\mycomment{\mbox{compute $S$ explanatory power~(Eq.~\ref{eq:item_power})}}}
}

\Return $isCF$, \textit{explPower}





\end{procedure}

 \stitle{Algorithm Description.}
Algorithm~\ref{algo:infuencerebuild} implements the
\linebreak
\mbox{\textit{ExpRebuild}} strategy.
The algorithm begins by computing the total score for all items and sorting them into a ranked list $H$ (\textit{lines}~1--5). It then incrementally builds a candidate set $S$ (\textit{lines}~6--11), invoking the procedure
\linebreak
\textsf{checkForCFandComputeExplPwr} (Proc.~2) at each step to determine whether a counterfactual explanation has been found and to compute the explanatory power of the current set.

Once a counterfactual explanation is identified, the items in $S$ are ordered according to their explanatory power (\textit{line}~14). The algorithm then reconstructs the explanation by incrementally adding items in descending explanatory power order (\textit{lines}~15--18), checking after each addition whether the counterfactual condition holds.

The first subset that satisfies the counterfactual condition is returned as the final explanation. If no such subset is found, the full set $S$ is returned (\textit{line}~19). If no counterfactual explanation is identified during the initial phase, the algorithm returns the empty set (\textit{line}~21).

\stitle{{Complexity: Time, Space, and Recommender Calls.}}

\vspace{-\bigskipamount}
\sititle{Time Complexity}:
In \textit{line} 3 item scores are computed for all items $i \in I$ using $d$ metric functions $f_1, \dots, f_d$.
As shown in Section~\ref{sec:pareto_algo}, the complexity to compute the metric functions for an item is $O(|U|  \phi)$. Thus, computing all item scores has a total cost:
$O\left( |I| |U| \phi \right)$.

Sorting the scored list $H$ costs $O(|I| \log |I|)$.
The main loop (\textit{lines}~7--11) adds items one-by-one to the candidate set $S$ and invokes the \textsf{checkForCFandComputeExplPwr} procedure at each step. In the worst case, the loop is performed $|I|$ times.


The \textsf{checkForCFandComputeExplPwr} procedure calls the recommender system once (\textit{line} 1), and computes the item explanatory  power score (Eq.~\ref{eq:item_power}), which can be done in $O(1)$ (if a hash-based indexing is used over $L$).
Hence, each invocation of \textsf{checkForCFandComputeExplPwr} costs: $O(\phi)$.

 Therefore, the total cost of the main loop (\textit{lines}~7--11) is: $O(|I| \phi)$.

Next, if a counterfactual explanation is found (\textit{line}~12), the algorithm sorts $S$ based on  item explanatory power scores, requiring $O(|I| \log |I|)$.
Then, in the  loop (\textit{lines}~15--18), the algorithm attempts to rebuild a more concise valid explanation set $P \subseteq S$ by adding one item at a time and calling \textsf{isCF}. In the worst case, $|S| = |I|$, so this phase invokes the recommender system at most $|I|$ times, resulting in $O(|I|\phi)$. 

  Summing up all parts, the overall \textit{worst-case time complexity} of the \textit{ExpRebuild} algorithm is:
\[
\begin{aligned}
O\big(& |I| |U|\phi + |I| \log |I| + |I|\phi  
      + |I| \log |I| + |I|\phi \big) \\
&= O\left( |I| |U|\phi + |I| \log |I|\right)
\end{aligned}
\]

\sititle{Space Complexity}:
The \textit{space complexity} is $O(|I|)$, for storing lists $H$, $S$, and $P$.

\sititle{Recommender Calls}:
In the worst case, the number of recommender calls for metrics computations is $O(|I||U|)$, during the grow phase (via \textsf{checkForCFandComputeExplPwr}) is $O(|I|)$, 
and $O(|I|)$  calls during the rebuild phase (via \textsf{isCF}). 
Therefore, in the worst case, the \textit{total number of recommender calls} is $O(|I||U|)$.

\begin{table*}[t]
\centering
\caption{Summary of the main experimental findings and trade-offs among the proposed methods.}
\label{tab:algo_empirical_comparison}

\small
\setlength{\tabcolsep}{6pt}
\renewcommand{\arraystretch}{1.2}

\begin{tabularx}{\textwidth}{@{}
>{\raggedright\arraybackslash}p{2.60cm}
>{\centering\arraybackslash}p{1.55cm}
>{\centering\arraybackslash}p{1.58cm}
>{\centering\arraybackslash}p{1.55cm}
>{\centering\arraybackslash}p{2.85cm}
>{\raggedright\arraybackslash}X
@{}}

\toprule
\textbf{Method} &
\textbf{Cost} &
\textbf{Size} &
\textbf{Fairness} &
\textbf{Interpretability} &
\textbf{Utility} \\
 \dlineB

\textbf{GreedyGrow}
\par{\scriptsize\color{black!60} Sec.~\ref{sec:greedy_grow}}
& Very Low
& Very Large
& Very Low
& High
& Lower \\
\rowcolor{gray!7}

\textbf{ExpRebuild}
\par{\scriptsize\color{black!60} Sec.~\ref{sec:influence_rebuild}}
& Low
& Moderate
& Low
& Very High\textsuperscript{a}
& Highest on MovieLens at $\alpha=0.3, 0.5$ \\

\textbf{Grow\&Prune}
\par{\scriptsize\color{black!60} Sec.~\ref{sec:grow_and_prune}}
& Moderate
& Small
& High
& Moderate
& Highest on MovieLens at $\alpha=0.7$ \\

\rowcolor{gray!7}
\textbf{FixedWindow}
\par{\scriptsize\color{black!60} Sec.~\ref{sec:fixed_window}}
& High
& Very Small
& Very High
& Low
& Highest on Amazon \\

\textbf{Pareto-filtering}\textsuperscript{*}
\par{\scriptsize\color{black!60} Sec.~\ref{sec:pareto_algo}}
& Mixed\textsuperscript{b}
& Mixed
& Mixed
& Mixed
& Mixed\textsuperscript{c} \\

\bline
\end{tabularx}

\vspace{4pt}

\begin{minipage}{\textwidth}
\footnotesize
\setlength{\parskip}{2pt}

\noindent\textsuperscript{*}
Pareto filtering is used as an optional preprocessing step
of each algorithm.

\noindent\textsuperscript{a}
\textit{ExpRebuild} leads in interpretability on MovieLens; \textit{GreedyGrow} and \textit{Grow\&Prune} have similar scores on Amazon.

\noindent\textsuperscript{b}
Filtering increases calls for all methods on MovieLens
Group~5, but reduces them for \textit{GreedyGrow}, \textit{Grow\&Prune},
and \textit{ExpRebuild} in Group~10 on both datasets.

\noindent\textsuperscript{c}
Filtering lowers utility on MovieLens but increases
it on Amazon.
\end{minipage}
\end{table*}

\begin{table}[h]
\centering
\small
\caption{Number of Users, Items, and Ratings}
\begin{tabular}{cccc}
\tline
\textbf{Dataset} & \textbf{Users} & \textbf{Items} & \textbf{Ratings} \\ \dlineB
\addlinespace[3pt]
\textbf{MovieLens} & 69,878 & 10,681  & 10,000,054 \\
\addlinespace[2pt]
\textbf{Amazon} & 344,747 &   373,665 & 5,573,065 \\
\addlinespace[1pt]
\bline
\end{tabular}
\label{tab:datasets}
\end{table}

\section{Experimental Evaluation}
\label{sec:exp}

We evaluate our methods on MovieLens 10M and Amazon,
examining explanation size, recommender-call cost,
interpretability, and fairness across group sizes.
We evaluate the impact of Pareto-filtering and we also compare our methods with an exhaustive search approach.
Finally, we assess the stability of the results through
bootstrap resampling and analyze utility under different
weights for minimality and interpretability.

Table~\ref{tab:algo_empirical_comparison} summarizes the main findings of the experimental evaluation across five key characteristics of the proposed methods: cost, explanation size, fairness, interpretability, and utility. These findings are further discussed in Section~\ref{sec:discussion}.

\subsection{Experimental Setting}
\label{sec:exp_setting}

\stitle{Datasets.} We employed two benchmark datasets, \textit{MovieLens}~10M \cite{Harper:2015:MDH:2866565.2827872}
and \textit{Amazon}~\cite{he2016ups}, which differ considerably in their characteristics. While MovieLens represents a dense and well-studied domain of movie ratings, the Amazon dataset is much sparser and reflects a more heterogeneous item space. Table~\ref{tab:datasets} presents the characteristics of each dataset.

\stitle{Evaluated Methods.}
We evaluate each of the proposed methods in two variants: with and without
Pareto filtering. The \textit{Pareto-filtering} variants apply the Pareto algorithm as a preprocessing step to reduce the candidate set.
The variant without filtering are denoted as \textit{Sorted List}.

We also compare our methods with an exhaustive search (\textit{Exhaustive}), which
examines item combinations in increasing order of cardinality until a valid counterfactual
explanation is identified.
We implemented all methods in Python 3.12.7, using NumPy, pandas, SciPy, and scikit-learn for numerical computations, data processing, and sparse matrix operations.

\stitle{Group Formation.}
To represent realistic group recommendation scenarios, we simulated groups of two sizes: five members (\textit{Group 5}), reflecting smaller and more common decision-making units, and ten members (\textit{Group 10}), representing larger and more diverse groups. We limited our analysis to users with a minimum of 50 ratings to ensure sufficient preference information and minimize noise in the recommendation process. This choice also increases the size of the candidate item pool, creating a more challenging environment for identifying group counterfactual explanations.
We randomly generate 50 groups of each size.

Furthermore, regarding the experiments where exhaustive search is used, additional groups are constructed.
Due to the high computational complexity of exhaustive search, evaluating it on arbitrary groups is impractical within a reasonable time.
To enable the comparisons with our methods, we selected 50 randomly generated groups of five users for which exhaustive search was able to identify a valid counterfactual explanation within a budget of 1,000 recommender calls.
We refer to this selected set as \textit{Group 5-Exst}.
Although this selection focuses on groups where Exhaustive method succeeds within the call
budget, it enables a direct comparison with our methods.

To assess the sensitivity of the results to the selected groups, we additionally perform
a bootstrap stability analysis.

\stitle{Evaluation Scenario.}
We report results averaged over the 50 randomly generated groups for each group size.
Each group recommendation list contains 10 items.
To ensure consistency and reduce randomness, we select
the top-ranked item in the original group recommendation
list as the target item.
An explanation is considered successful
if removing the items in the explanation causes the target
to disappear from the top-10 recommendation list.
We use a budget of 1,000 recommender calls and set the
initial window size to $w=15$ for the \textit{FixedWindow}
algorithm.

\stitle{Evaluated Metrics.}
We evaluate the methods in terms of (a)~explanation size;
(b)~cost (number of recommender calls); (c)~interpretability;
(d)~fairness; and (e)~utility. We further assess the stability
of the results using bootstrap confidence intervals.

\stitle{Recommender System.}
Group recommendation lists were generated through a two-step process. First, individual recommendation scores were predicted for each group member using user-based collaborative filtering with cosine similarity. The recommender considered all other users without restricting the neighborhood to a fixed number of users. It computed predictions using similarity-weighted ratings, assigning a score of zero when no similarity information was available. The predicted scores were normalized to a common range of $[0,5]$, and items already rated by each user were excluded from their individual recommendations.

Second, group-level scores were computed by averaging the predicted scores across all group members. Items previously rated by any group member were excluded from the final candidate set, and the 10 highest-scoring items were selected to form the group recommendation list.

For counterfactual evaluation, we removed the corresponding items from group members' rating profiles and recomputed their individual predictions and group recommendations. This configuration provides a consistent recommendation setting for evaluating our counterfactual explanation methods.

\begin{figure*}[t]
    \centering
    \begin{subfigure}[t]{0.49\linewidth}
        \centering
        \includegraphics[width=\linewidth]{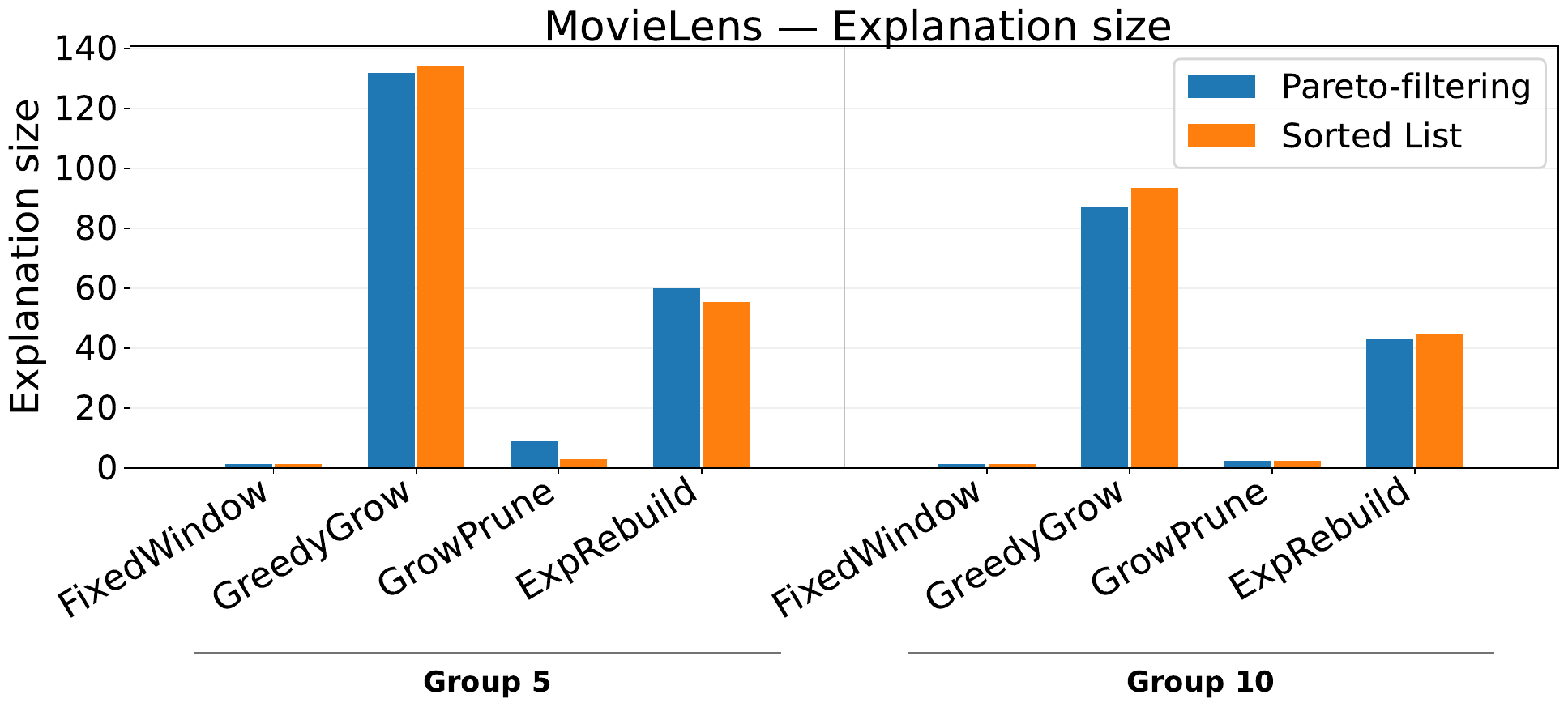}
        \caption{Explanation size}
        \label{fig:movielens_size}
    \end{subfigure}
    \hfill
    \begin{subfigure}[t]{0.49\linewidth}
        \centering
        \includegraphics[width=\linewidth]{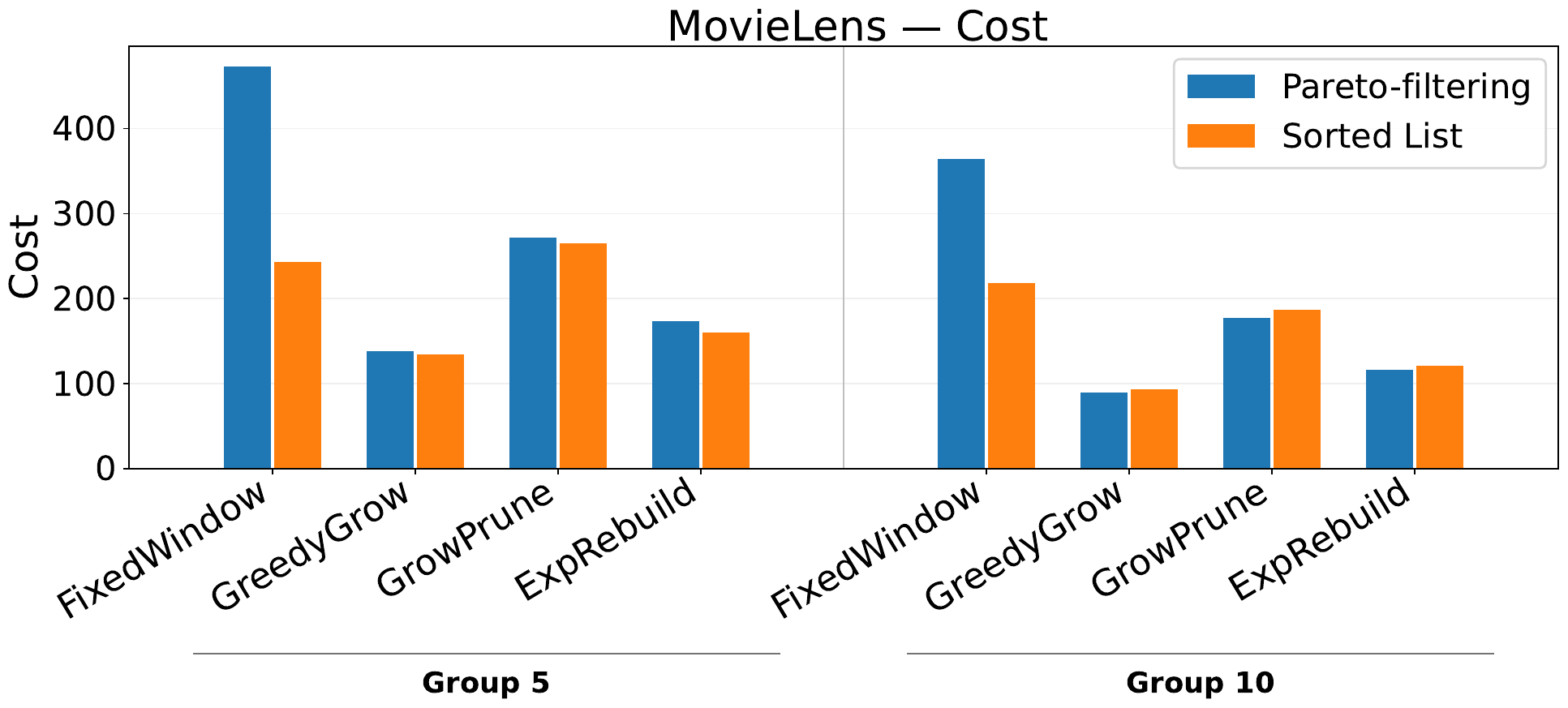}
        \caption{Cost}
        \label{fig:movielens_cost}
    \end{subfigure}
    \caption{Explanation size and Cost for MovieLens (Group 5 \& Group 10).}
    \label{fig:movielens_size_cost}
\end{figure*}

\begin{figure*}[t]
    \centering
    \begin{subfigure}[t]{0.49\linewidth}
        \centering
        \includegraphics[width=\linewidth]{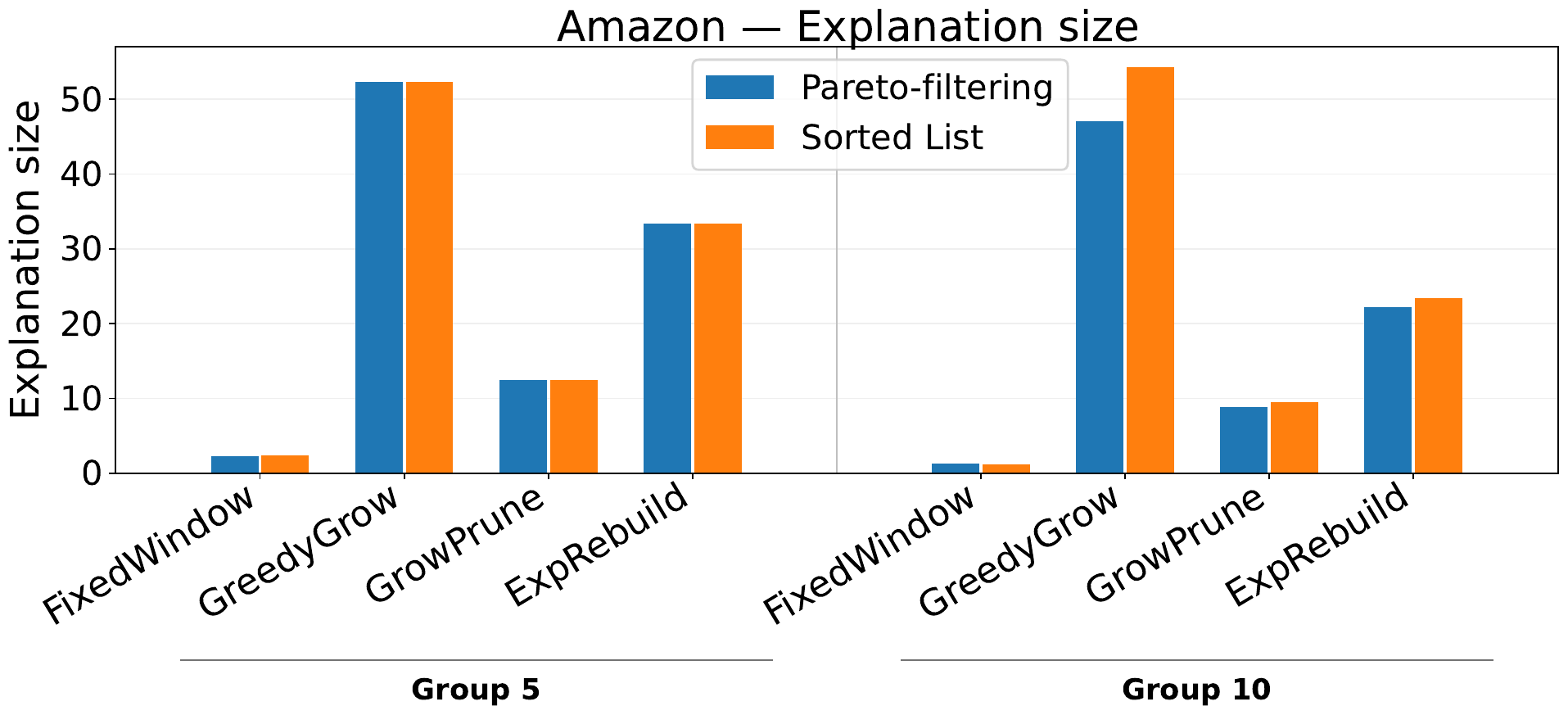}
        \caption{Explanation size}
        \label{fig:amazon_size}
    \end{subfigure}
    \hfill
    \begin{subfigure}[t]{0.49\linewidth}
        \centering
        \includegraphics[width=\linewidth]{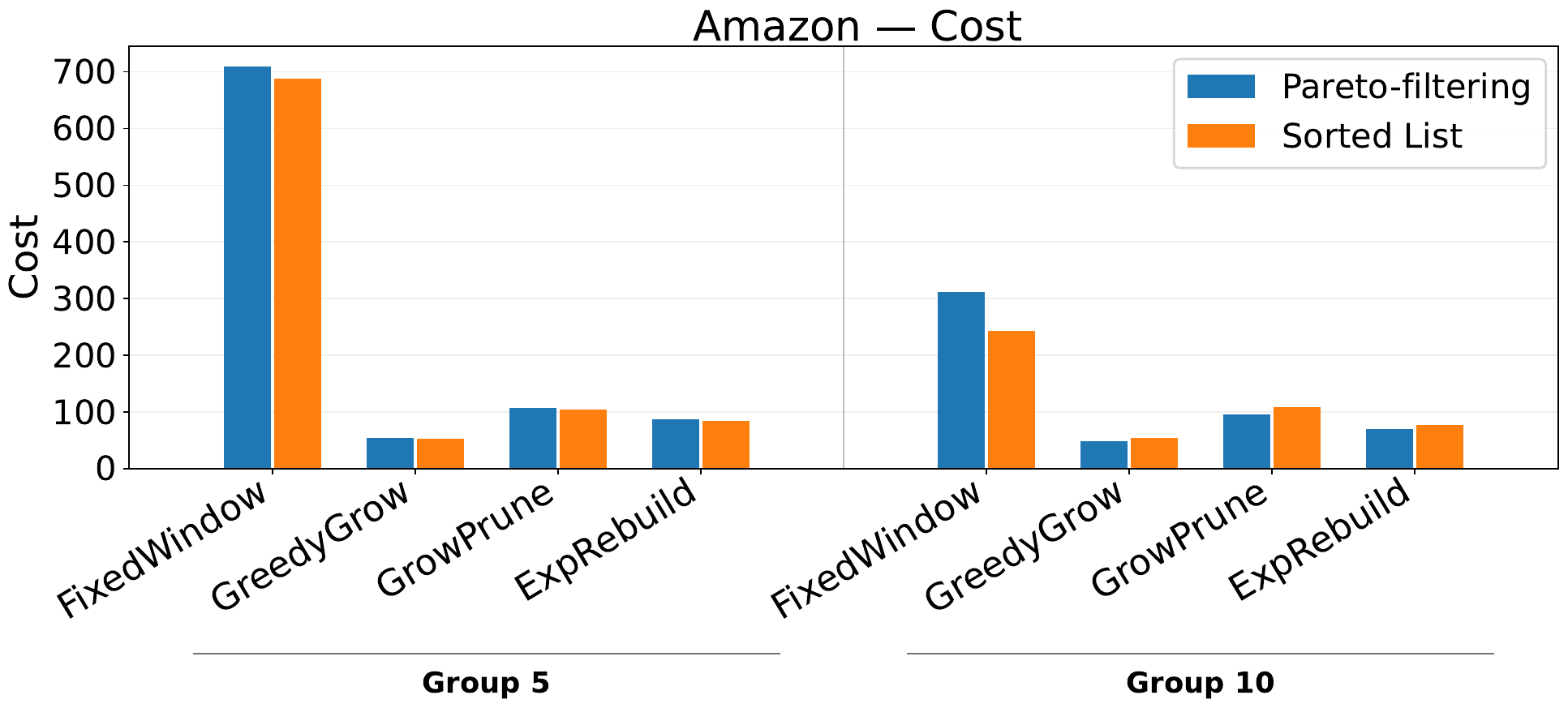}
        \caption{Cost}
        \label{fig:amazon_cost}
    \end{subfigure}
    \caption{Explanation size and Cost for Amazon dataset (Group 5 \& Group 10).}
    \label{fig:amazon_size_cost}
    \vspace{-8pt}
\end{figure*}

\subsection{Explanation Size and Cost}

\stitle{MovieLens.}
Figure~\ref{fig:movielens_size_cost} reports the explanation size and cost for the MovieLens dataset, comparing groups of sizes 5 and 10 with and without Pareto-filtering.

For groups of 5, \mbox{GreedyGrow} requires the fewest calls in both variants but produces the largest explanations. With Sorted List, it requires $134$ calls on average, compared with 265 for Grow\&Prune and 243 for \mbox{FixedWindow}. However, its explanations contain 134 items on average, compared with 2.9 for Grow\&Prune and 1.3 for \mbox{FixedWindow}. \mbox{ExpRebuild} provides an intermediate alternative, requiring approximately 160 calls and producing explanations of 55 items. With Pareto-filtering, Grow\&Prune explanation size increases from 2.9 to 9.3 items (see also Sect.~\ref{sec:paretoeval}).

For groups of 10, Grow\&Prune and \mbox{FixedWindow} again produce compact explanations, while \mbox{GreedyGrow} remains the least expensive. In particular, Grow\&Prune returns 2.41 items with Sorted List and 2.48 with Pareto-filtering, at a substantially lower call cost than \mbox{FixedWindow}.

\stitle{Amazon.}
Figure~\ref{fig:amazon_size_cost} shows explanation size and cost for Amazon.
For groups of 5, \mbox{GreedyGrow} requires the fewest calls (approximately 53) but produces the largest explanations (about 52 items).
With Sorted List, Grow\&Prune reduces the average explanation size to about 12 items, at a cost of 105 calls.
ExpRebuild offers an intermediate alternative, producing 33 items with 84 calls.
\mbox{FixedWindow} returns the shortest explanations (2.4 items), but at a much higher cost: approximately 688 calls with Sorted List and 709 with Pareto-filtering.
Its local refinement becomes expensive when a suitable counterfactual window is difficult to locate.

For groups of 10, the same trade-off is observed. \linebreak \mbox{GreedyGrow} requires the fewest calls, while \mbox{FixedWindow} produces the shortest explanations but has the highest cost. Grow\&Prune and ExpRebuild offer intermediate alternatives.

\stitle{Summary.}
 Overall, \mbox{GreedyGrow} requires the fewest recommender calls but produces the largest explanations, whereas \mbox{FixedWindow} typically returns the shortest explanations at a substantially higher cost. Grow\&Prune offers a favorable compromise between explanation size and cost, while ExpRebuild provides an intermediate alternative.

\subsection{Methods with Pareto-filtering}
\label{sec:paretoeval}
Figures~\ref{fig:movielens_size_cost} and~\ref{fig:amazon_size_cost} compare the Sorted List and Pareto-filtering variants for both group sizes. Pareto-filtering does not provide a uniform reduction in call cost.
For MovieLens (Fig.~\ref{fig:movielens_size_cost}) in Group 5, it increases the cost of all four heuristics and raises the mean explanation size of Grow\&Prune.
For {Group 10}, it reduces the call cost of \mbox{GreedyGrow}, Grow\&Prune, and ExpRebuild, while their explanation sizes change less markedly.

For Amazon (Fig.~\ref{fig:amazon_size_cost}) in Group 5, the two variants produce similar explanation sizes and call costs.
For {Group 10}, Pareto-filtering reduces both measures for \mbox{GreedyGrow}, Grow\&Prune, and ExpRebuild.

\vspace{-4pt}
\stitle{Summary.}
Overall, Pareto-filtering offers no consistent advantage across datasets and group sizes. It reduces cost for several algorithms in Group~10, particularly on Amazon, but may increase cost or explanation size in MovieLens Group~5.

 \begin{figure*}
    \centering
    \begin{subfigure}[t]{0.49\linewidth}
        \centering
        \includegraphics[width=\linewidth]{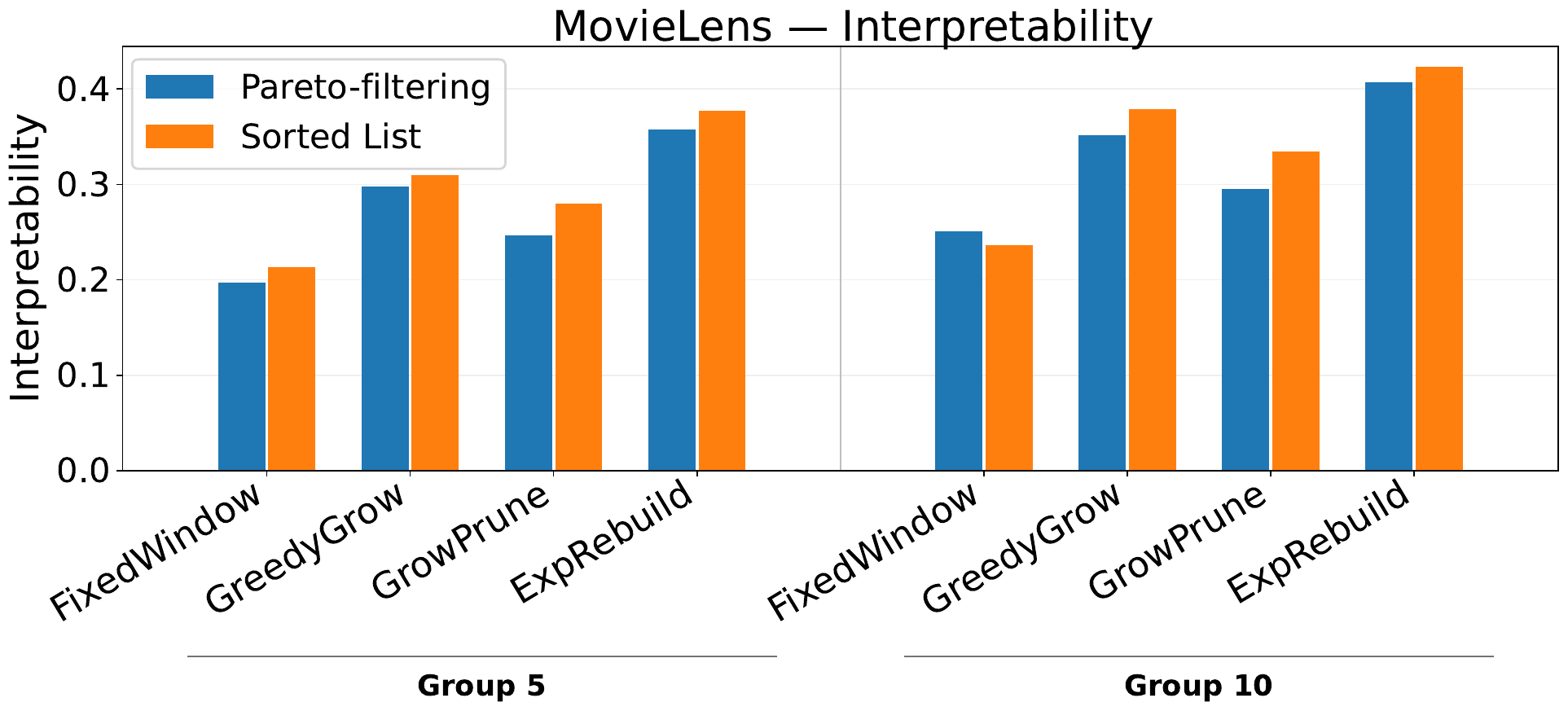}
        \caption{MovieLens}
        \label{fig:movielens_inter}
    \end{subfigure}
    \hfill
    \begin{subfigure}[t]{0.49\linewidth}
        \centering
        \includegraphics[width=\linewidth]{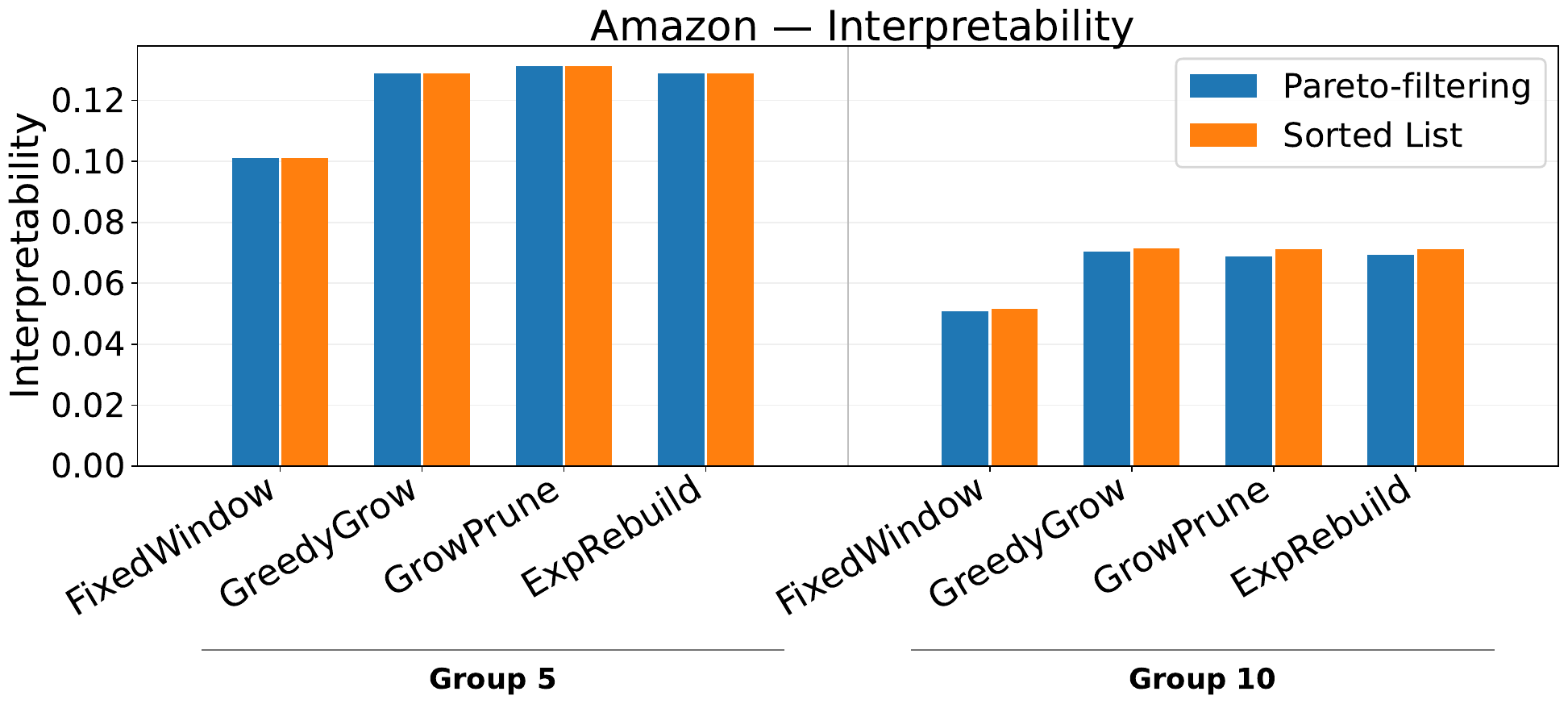}
        \caption{Amazon}
        \label{fig:amazon_inter}
    \end{subfigure}
    \caption{Explanation Interpretability for MovieLens and Amazon (Group 5 and 10).}
    \label{fig:interpretability}
\end{figure*}

\begin{figure*}
    \centering
    \begin{subfigure}[t]{0.49\linewidth}
        \centering
        \includegraphics[width=\linewidth]{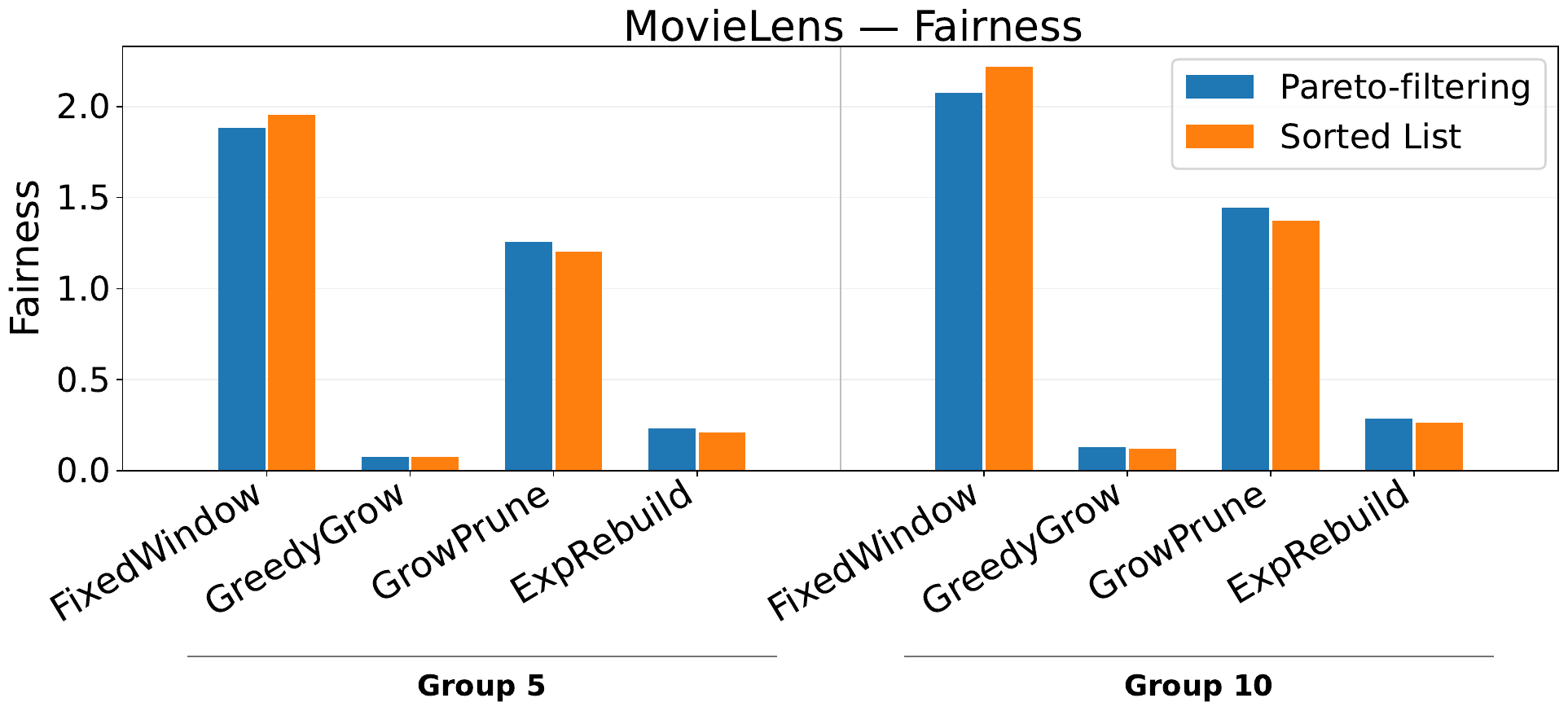}
        \caption{MovieLens}
        \label{fig:movielens_fairness}
    \end{subfigure}
    \hfill
    \begin{subfigure}[t]{0.49\linewidth}
        \centering
        \includegraphics[width=\linewidth]{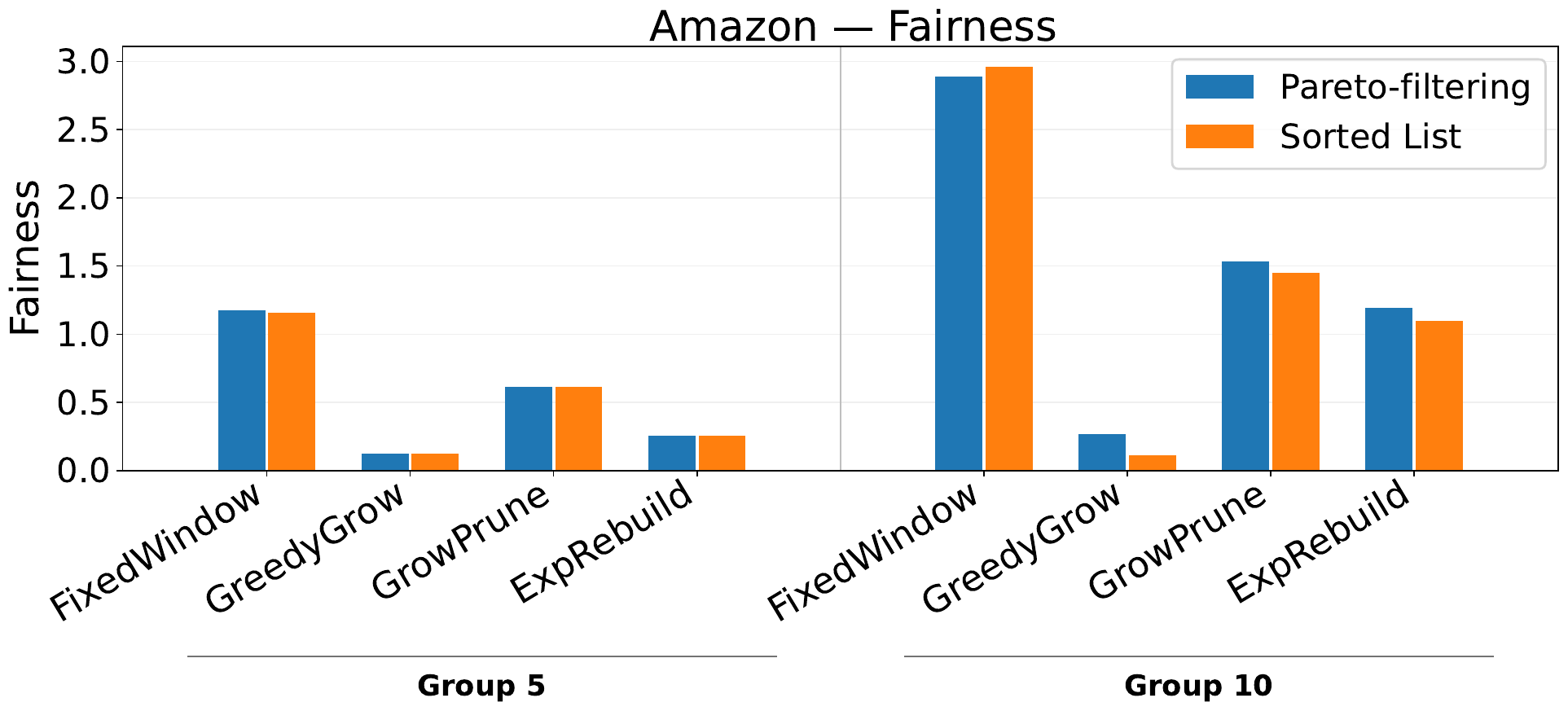}
        \caption{Amazon}
        \label{fig:amazon_fairness}
    \end{subfigure}
    \caption{Explanation fairness for MovieLens and Amazon  (Group 5 and 10).}
    \label{fig:fairness}
    \vspace{-8pt}
\end{figure*}

\subsection{Explanation Interpretability}

We evaluate recognition-based interpretability using Eq.~\ref{eq:recognition}, which combines item recognition within the group and among users outside it.
Figures~\ref{fig:movielens_inter} and~\ref{fig:amazon_inter} show different patterns across the two datasets.
On MovieLens (Fig.~\ref{fig:movielens_inter}), ExpRebuild obtains the highest score in both group-size configurations and variants, with a score around 0.36--0.42.
\mbox{GreedyGrow} also returns relatively recognizable items, while Grow\&Prune and \mbox{FixedWindow}
return shorter explanations with lower recognition scores.  This indicates that compact explanations are not necessarily the most recognizable to group members.

On Amazon (Fig.~\ref{fig:amazon_inter}), interpretability scores are considerably lower, consistent with the dataset's sparser interactions.
For groups of size 5, \mbox{GreedyGrow}, Grow\&Prune, and ExpRebuild achieve similar interpretability, around 0.13, while \mbox{FixedWindow} is closer to 0.10.
For groups of size 10, the first three methods again obtain comparable scores, around 0.07. The difference between group sizes is more pronounced on Amazon than on MovieLens.

Pareto-filtering does not consistently improve interpretability. In some configurations, the Sorted List variant produces explanations containing more familiar items.
ExpRebuild is particularly competitive on MovieLens, whereas the three grow-based methods perform similarly on Amazon.

\vspace{-4pt}
\stitle{Summary.}
Overall, ExpRebuild achieves the highest interpretability on MovieLens, whereas \mbox{GreedyGrow}, \linebreak
Grow\&Prune, and ExpRebuild obtain similar scores on Amazon. Pareto-filtering does not consistently improve interpretability.

\subsection{Explanation Fairness Analysis}

Figure~\ref{fig:fairness} presents the fairness scores (Eq.~\ref{eq:fair}) for the MovieLens and Amazon datasets.
For MovieLens, \linebreak
\mbox{FixedWindow} obtains the highest scores, followed by \linebreak Grow\&Prune. For groups of 5 members, their scores are approximately 1.9 and 1.2, respectively, compared with less than 0.24 for ExpRebuild and 0.08 for \mbox{GreedyGrow}. The same ordering holds for groups of 10 members.
A similar pattern is observed for Amazon. \mbox{FixedWindow} and Grow\&Prune obtain higher fairness scores than ExpRebuild and \mbox{GreedyGrow}.
In particular, for groups of size 10, \mbox{FixedWindow} achieves fairness scores of approximately 2.9, while Grow\&Prune obtains approximately 1.5.

Since the fairness metric is based on the distribution of absolute item counts among group members, fairness scores should be interpreted alongside explanation size.

\vspace{-4pt}
\stitle{Summary.}
Overall, \mbox{FixedWindow} achieves the highest fairness scores, followed by Grow\&Prune, across both datasets and group sizes.

\begin{figure*}
    \centering
    \begin{subfigure}[t]{0.47\linewidth}
        \centering
        \includegraphics[width=\linewidth]{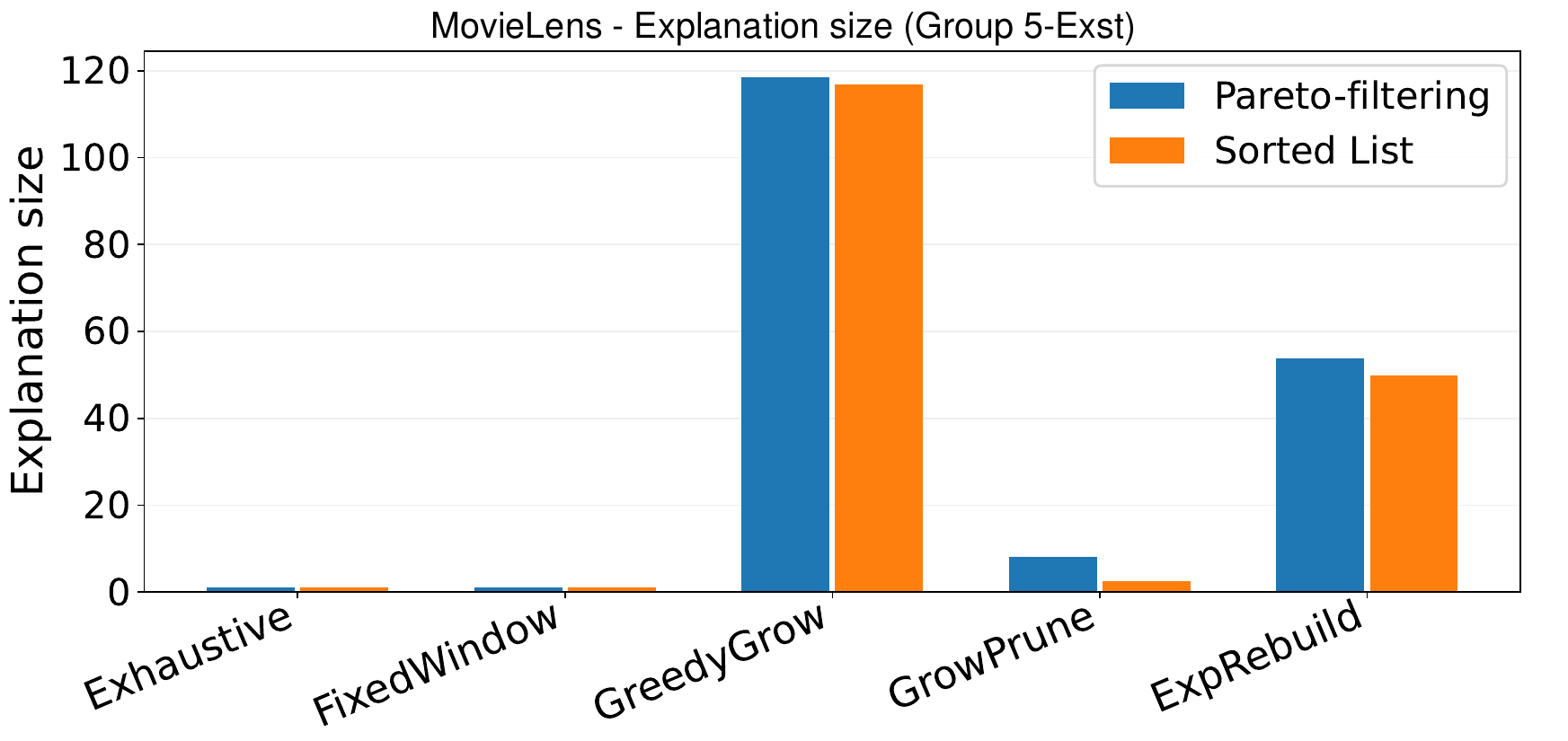}
        \caption{Explanation size}
        \label{fig:movielens_success_size}
    \end{subfigure}
    \hfill
    \begin{subfigure}[t]{0.47\linewidth}
        \centering
        \includegraphics[width=\linewidth]{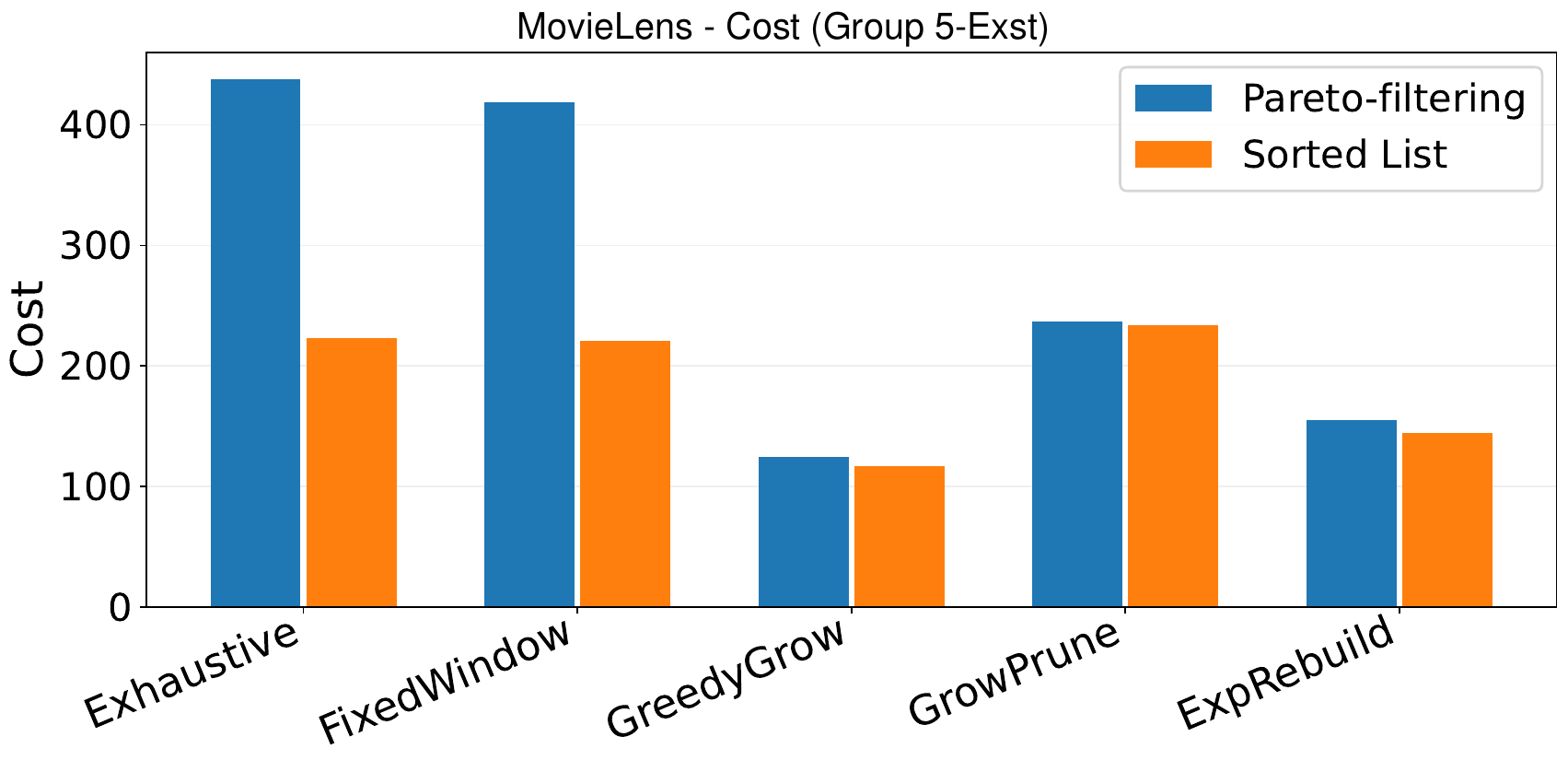}
        \caption{Cost}
        \label{fig:movielens_success_cost}
    \end{subfigure}
\caption{Explanation size and Cost on MovieLens for the \mbox{Exhaustive} Search setting ({Group 5-Exst}).}    \label{fig:movielens_success}
\end{figure*}

\begin{figure*}
    \centering
    \begin{subfigure}[t]{0.47\linewidth}
        \centering
        \includegraphics[width=\linewidth]{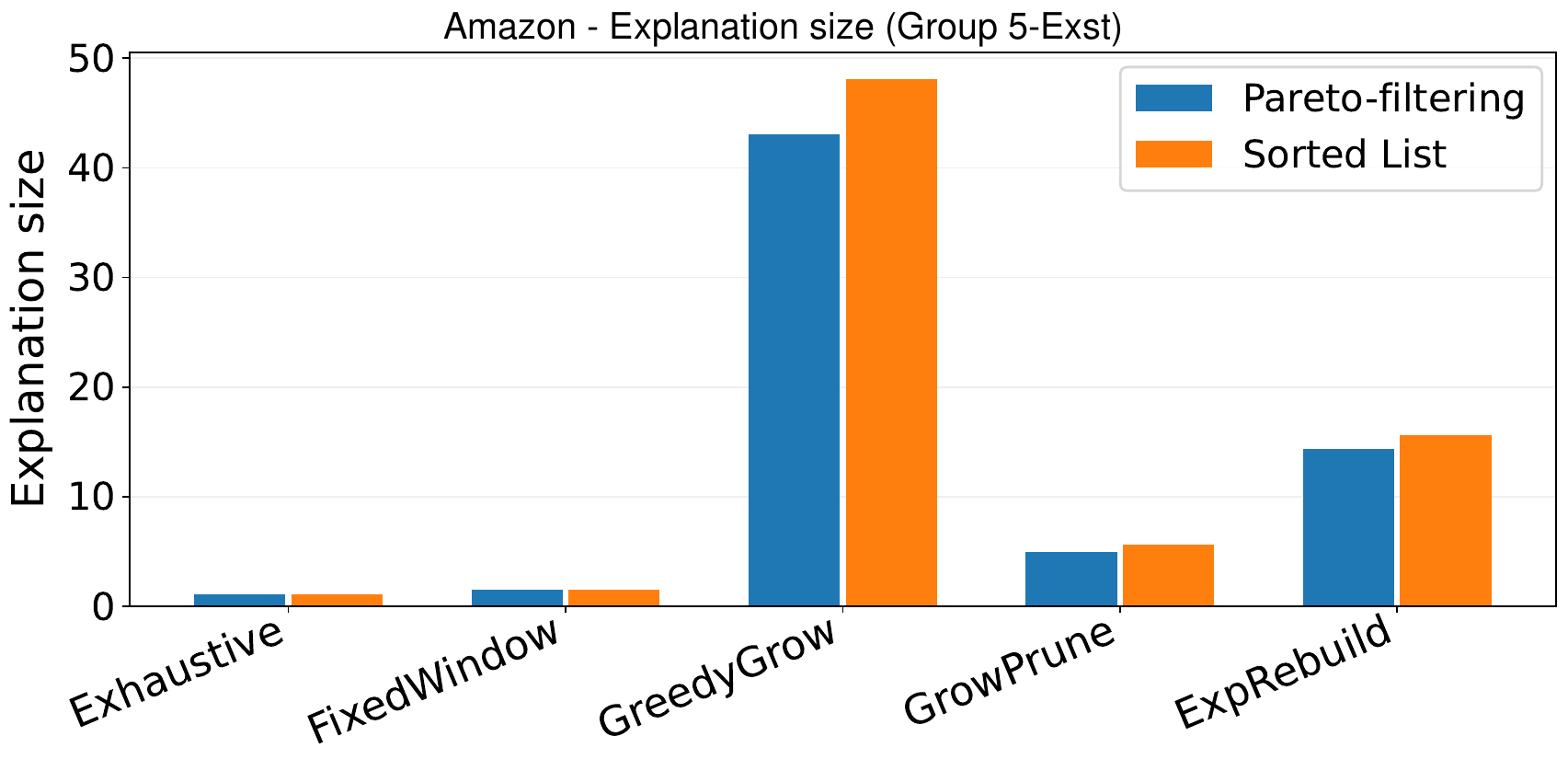}
        \caption{Explanation size}
        \label{fig:amazon_success_size}
    \end{subfigure}
    \hfill
    \begin{subfigure}[t]{0.47\linewidth}
        \centering
        \includegraphics[width=\linewidth]{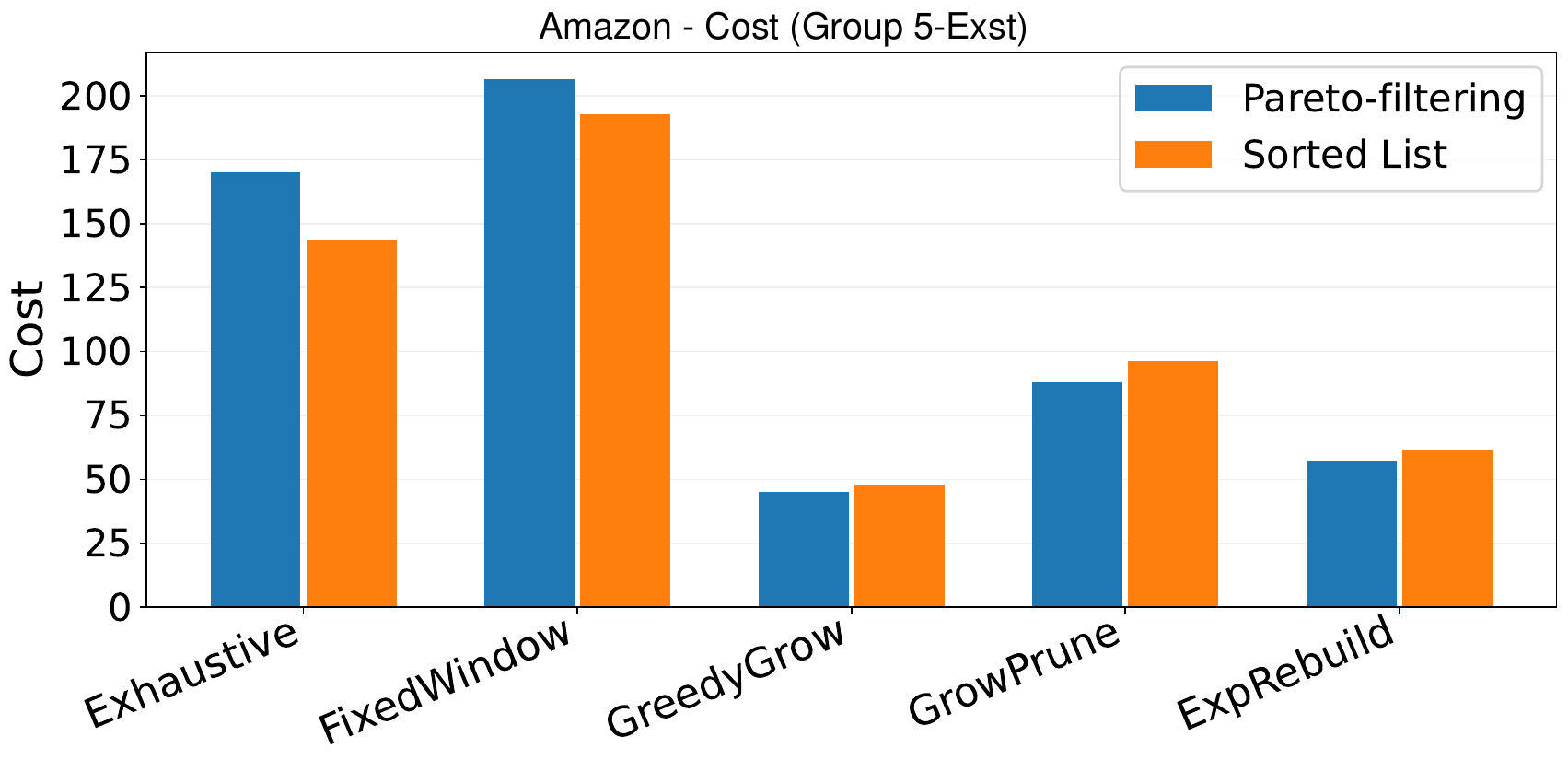}
        \caption{Cost}
        \label{fig:amazon_success_cost}
    \end{subfigure}
\caption{Explanation size and Cost on Amazon for the \mbox{Exhaustive} Search setting ({Group 5-Exst}).}
    \label{fig:amazon_success}
    \vspace{-8pt}
\end{figure*}

\subsection{Comparison with \mbox{Exhaustive} Search}
We compare the proposed algorithms with the \mbox{Exhaustive} approach on the groups of 5 members (Group 5-Exst) defined in the experimental setting.

Figure~\ref{fig:movielens_success} presents the MovieLens results.
With Sorted List, \mbox{Exhaustive}  returns explanations of approximately 1.07 items at 223 calls.
\mbox{FixedWindow} returns 1.14 items at a similar cost of 221 calls.
Grow\&Prune returns approximately 2.6 items at 234 calls.
\mbox{GreedyGrow} and ExpRebuild use approximately 117 and 144 calls, respectively, while returning larger explanations.
The comparison illustrates the different strengths of the proposed methods: \mbox{FixedWindow} approaches the compactness of \mbox{Exhaustive}, whereas \mbox{GreedyGrow} and ExpRebuild reduce calls at the expense of explanation size.

For Amazon (Fig.~\ref{fig:amazon_success}), \mbox{Exhaustive}  returns approximately 1.13 items at 144 calls with Sorted List.
\mbox{\mbox{FixedWindow}} produces similarly short explanations (1.5 items), although at a higher cost.
Grow\&Prune reduces the mean size from 48.1 items for \mbox{GreedyGrow} to 5.6 items, using approximately 96 calls.
ExpRebuild returns 15.6 items at about 62 calls, while \mbox{GreedyGrow} requires 48 calls.
In general, on Amazon, these three heuristics use fewer calls than \mbox{Exhaustive}, while \mbox{FixedWindow} uses more.

\vspace{-4pt}
\stitle{Summary.}
 Grow\&Prune is particularly effective when compactness must be balanced against computational cost. \linebreak
\mbox{FixedWindow}  returns small explanations in both datasets, with average sizes close to those of \mbox{Exhaustive}.
This is consistent with its local exhaustive refinement, which searches for smaller counterfactual subsets within a selected window.

\subsection{Explanation Utility}
To evaluate the utility of the generated explanations (Eq.~\ref{eq:expl_util}), we employ a weighted sum of minimality and interpretability, defined as $\alpha \, \fnc{min}(E)+ (1-\alpha)\,\fnc{interpr}(E)$, where $\alpha \in [0,1]$.
The weight $\alpha$ determines the relative importance of the two metrics. We consider three configurations, ${\alpha=0.3, 0.5, 0.7}$, and also report results for the \mbox{Exhaustive}.

\begin{figure*}[t]
    \centering

    \begin{subfigure}[t]{0.47\textwidth}
        \centering
        \includegraphics[width=\linewidth]{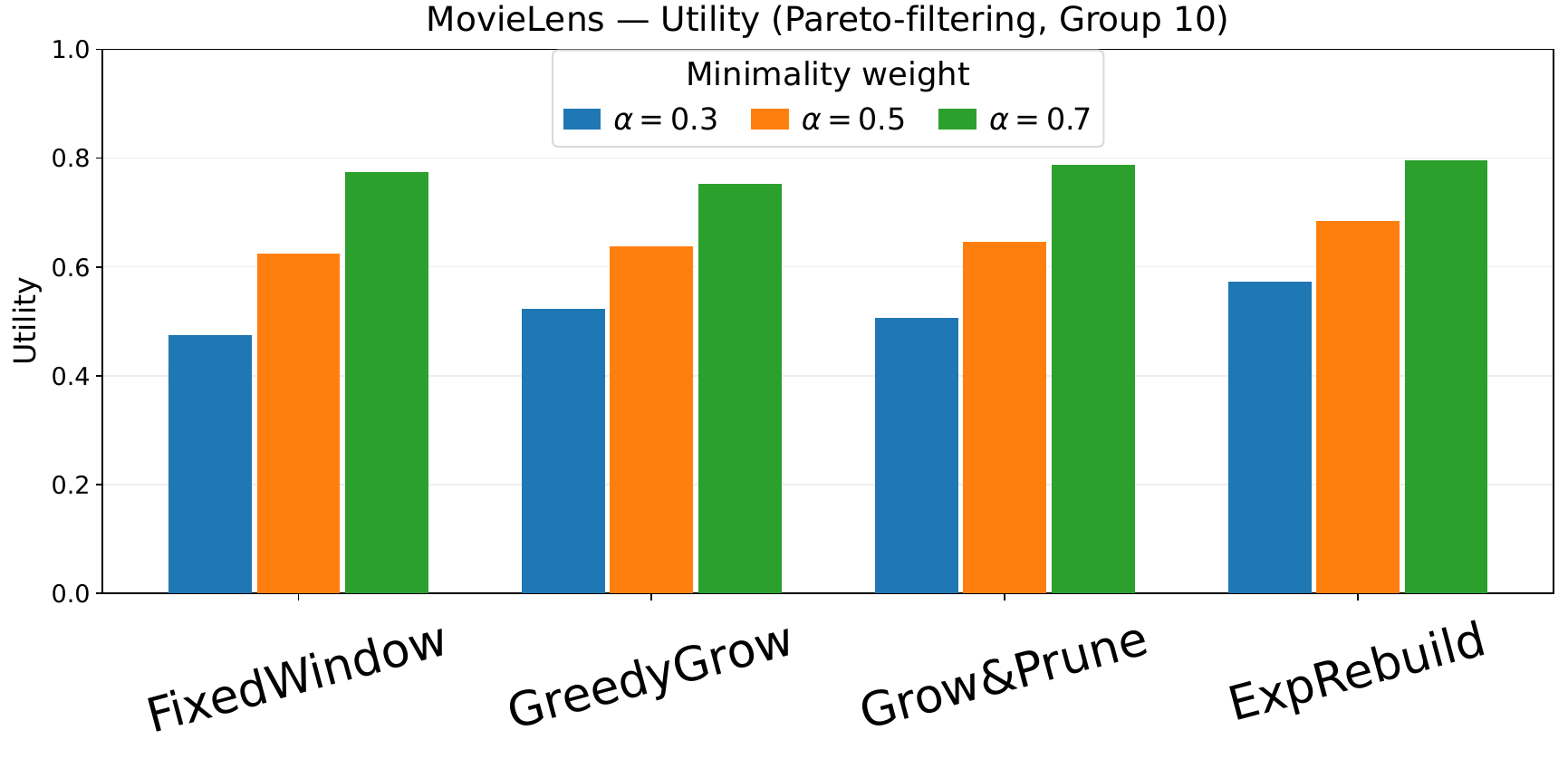}
        \caption{MovieLens (Pareto-filtering)}
        \label{fig:utility_ml_pareto_10}
    \end{subfigure}
    \hfill
    \begin{subfigure}[t]{0.47\textwidth}
        \centering
        \includegraphics[width=\linewidth]{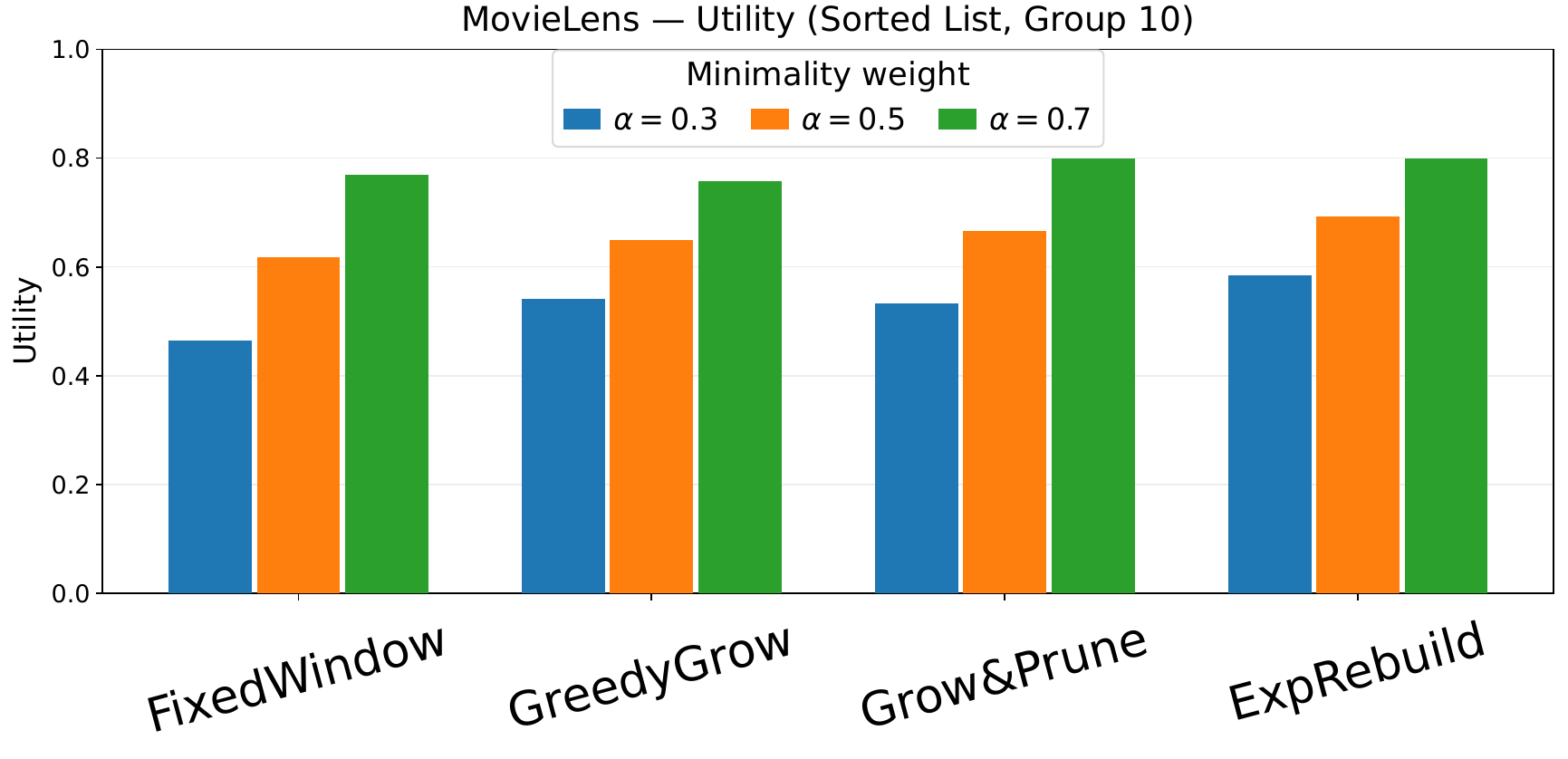}
        \caption{MovieLens  (Sorted List)}
        \label{fig:utility_ml_sorted_10}
    \end{subfigure}

    \vspace{0.5em}

    \begin{subfigure}[t]{0.46\textwidth}
        \centering
        \includegraphics[width=\linewidth]{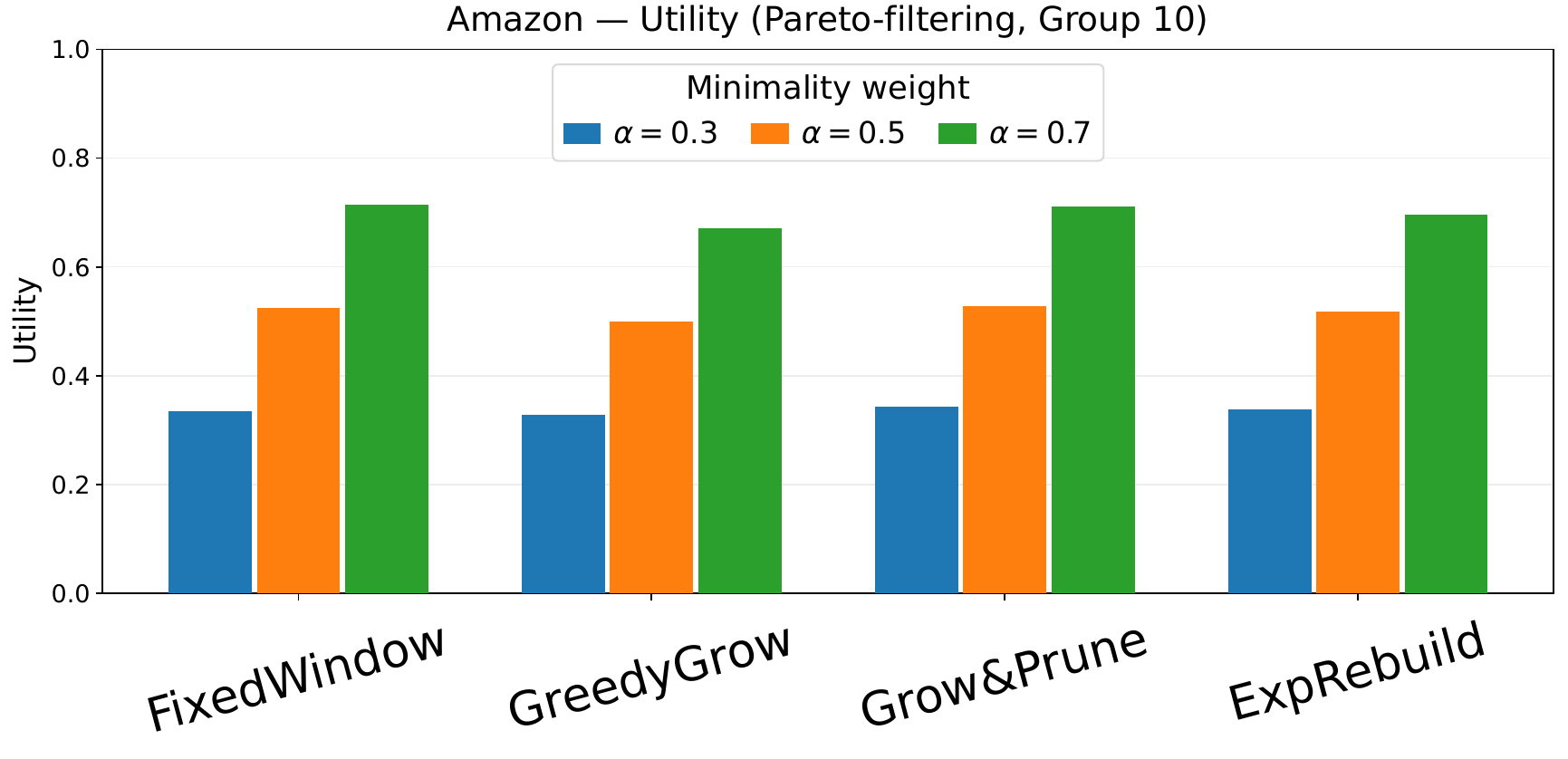}
        \caption{Amazon (Pareto-filtering)}
        \label{fig:utility_amazon_pareto_10}
    \end{subfigure}
    \hfill
    \begin{subfigure}[t]{0.46\textwidth}
        \centering
        \includegraphics[width=\linewidth]{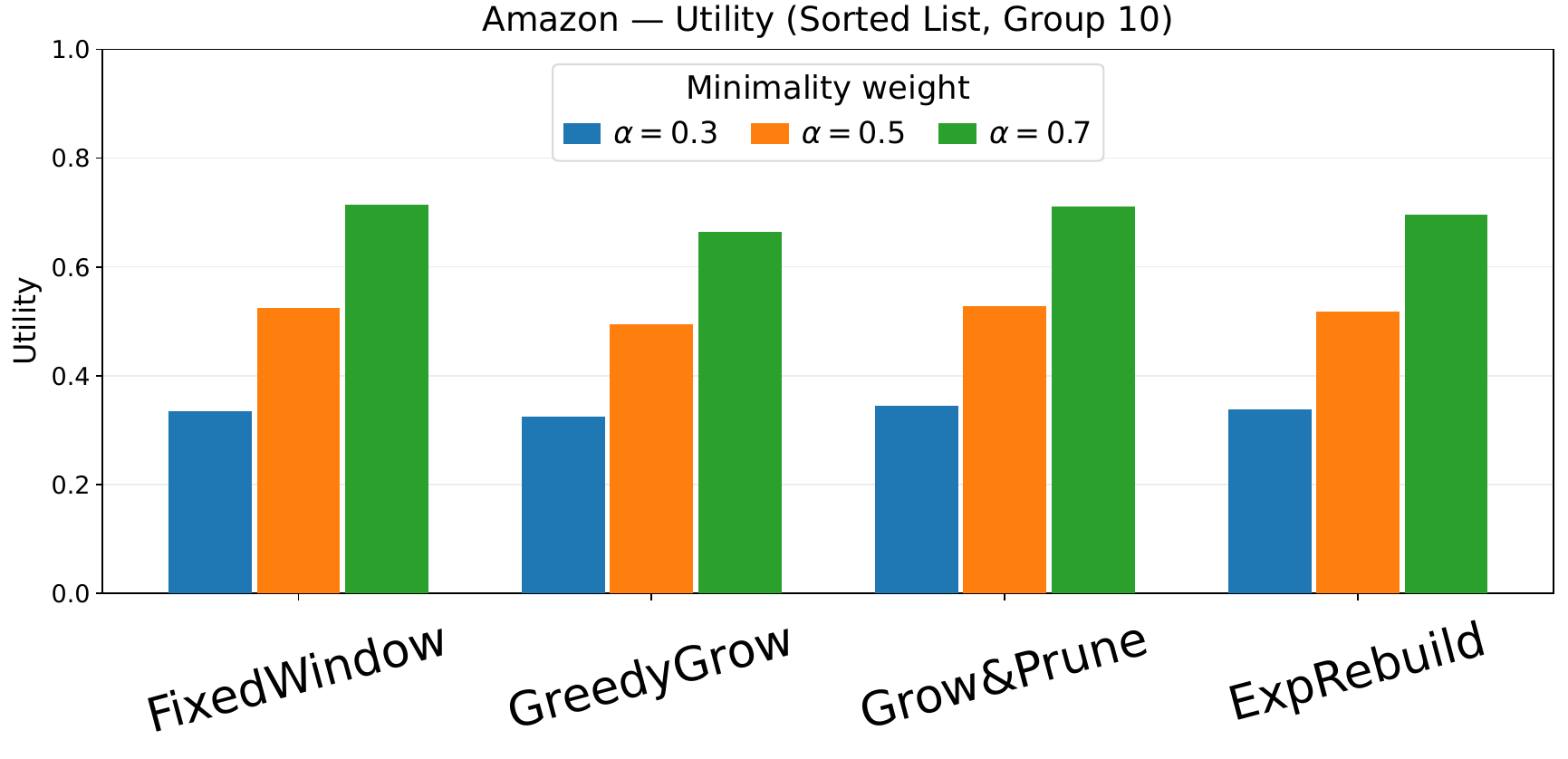}
        \caption{Amazon (Sorted List)}
        \label{fig:utility_amazon_sorted_10}
    \end{subfigure}

    \caption{Explanation utility for MovieLens and Amazon under varying weights for minimality and interpretability (Group 10).}
    \label{fig:utility_comparison}
    \vspace{-8pt}
\end{figure*}

Figure~\ref{fig:utility_comparison} shows that the preferred method depends on the dataset and the weight.
On MovieLens, ExpRebuild has the highest mean utility at weights 0.3 and 0.5, while Grow\&Prune leads at 0.7.
This ordering holds for both Sorted List and Pareto-filtering.  \mbox{FixedWindow} returns shorter explanations, but does not obtain the highest utility at any of the tested weights on MovieLens.
On Amazon, \mbox{FixedWindow} is the best-performing heuristic and remains very close to \mbox{Exhaustive}, with mean utility differences below 0.001; Grow\&Prune follows.

\begin{figure*}[t]
    \centering

    \begin{subfigure}[t]{0.48\textwidth}
        \centering
        \includegraphics[width=\linewidth]{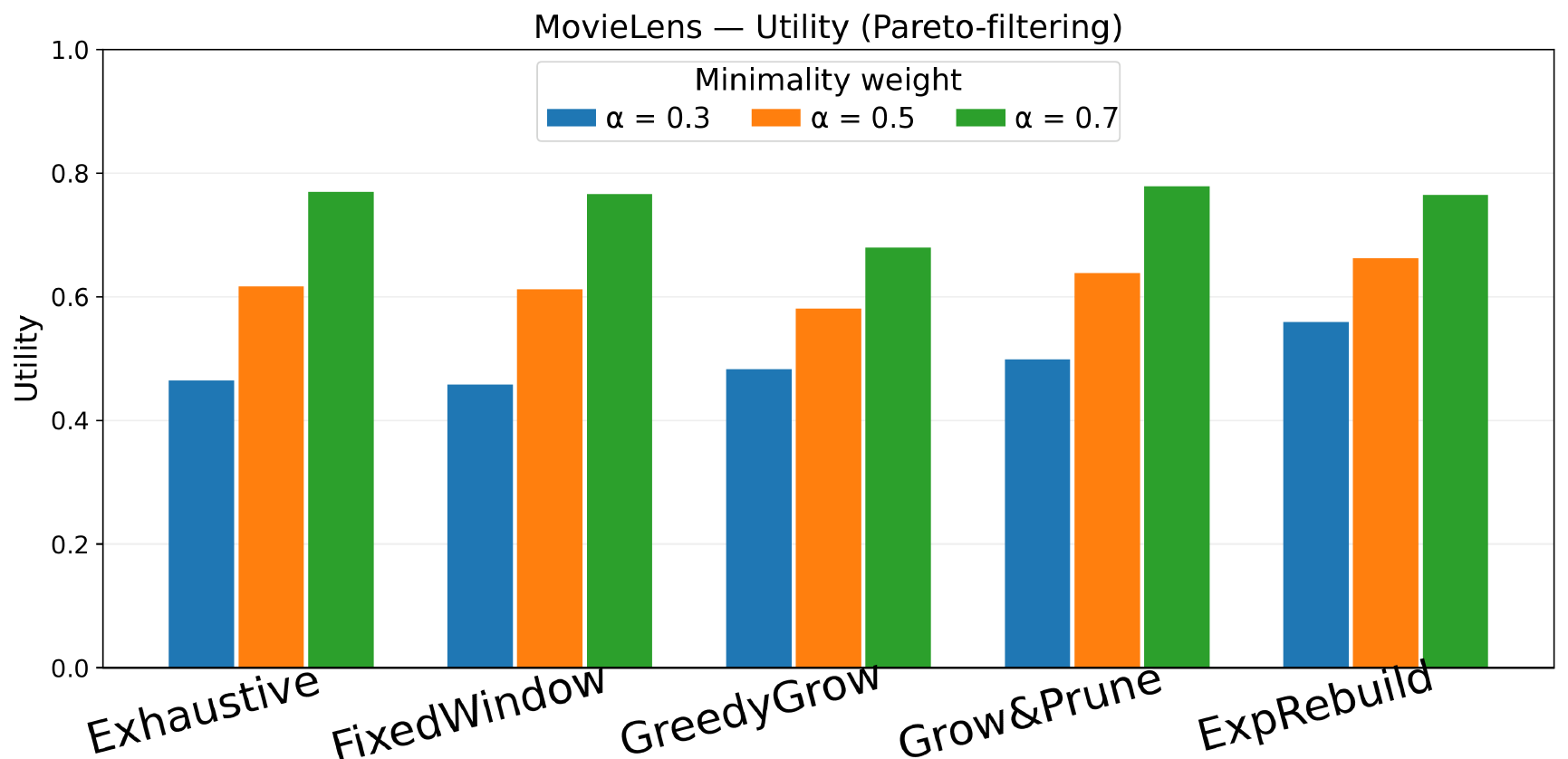}
        \caption{MovieLens (Pareto-filtering)}
        \label{fig:utility_ml_pareto}
    \end{subfigure}
    \hfill
    \begin{subfigure}[t]{0.48\textwidth}
        \centering
        \includegraphics[width=\linewidth]{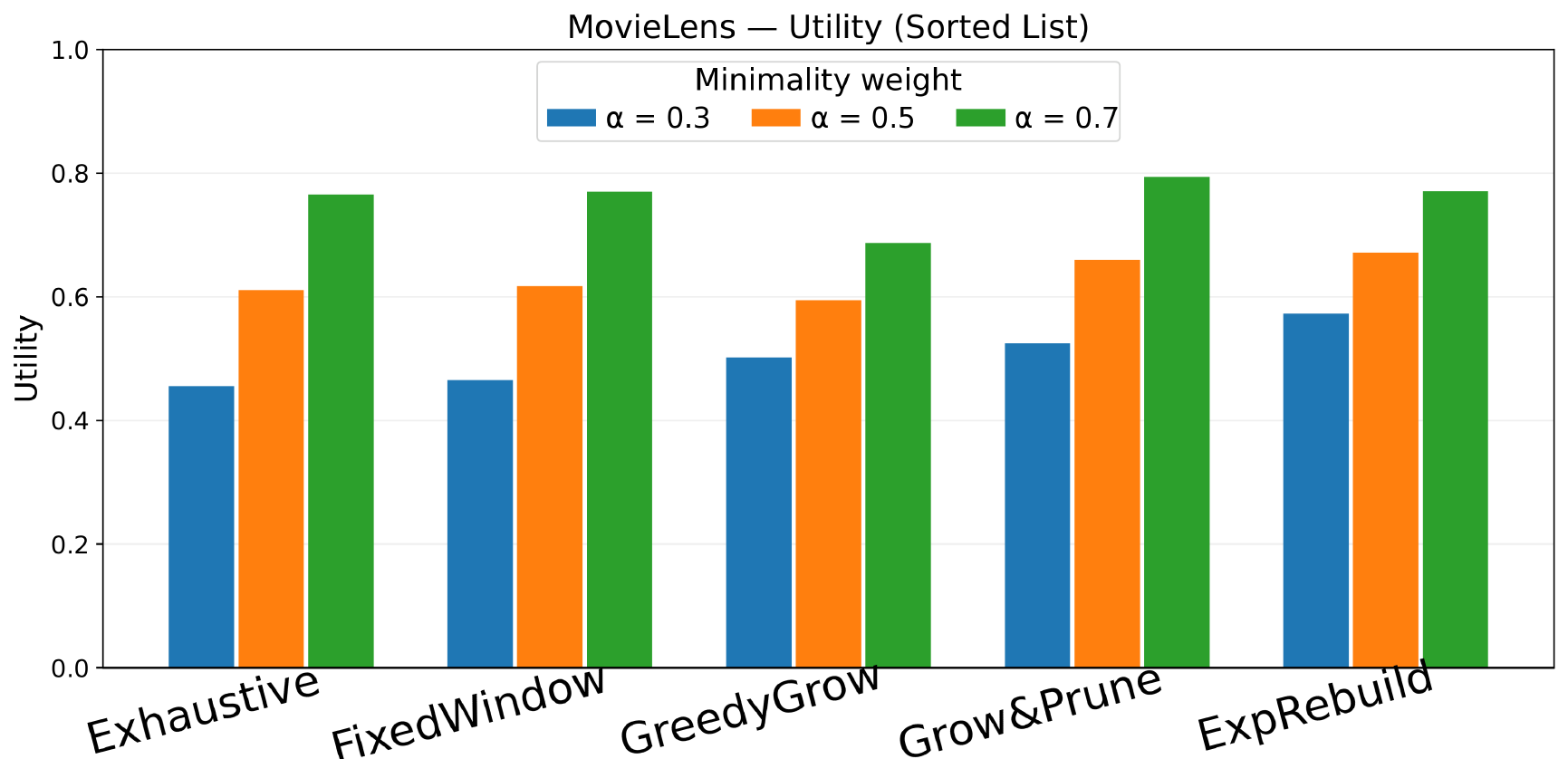}
        \caption{MovieLens  (Sorted List)}
        \label{fig:utility_ml_sorted}
    \end{subfigure}

    \vspace{0.5em}

    \begin{subfigure}[t]{0.48\textwidth}
        \centering
        \includegraphics[width=\linewidth]{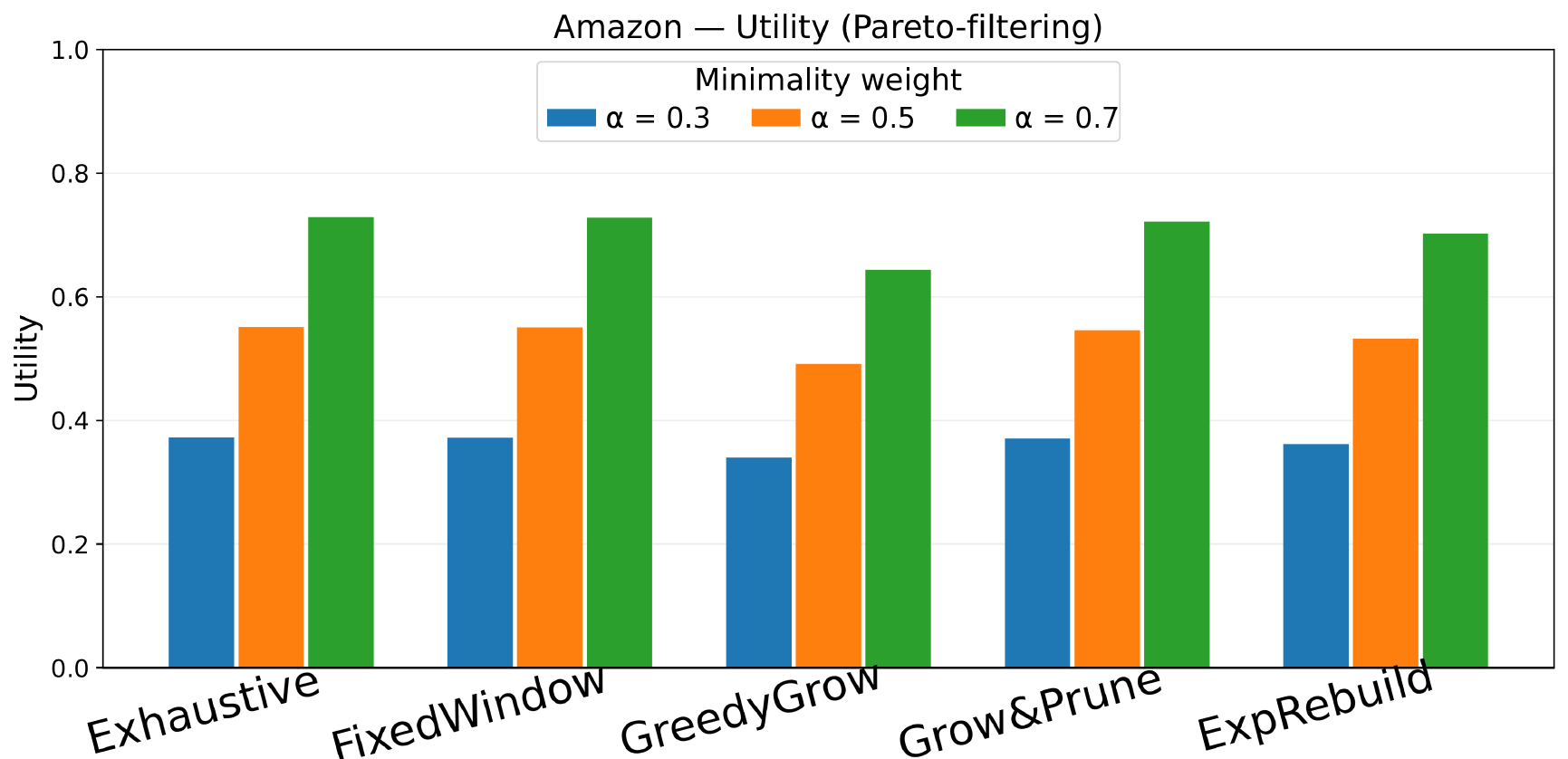}
        \caption{Amazon (Pareto-filtering)}
        \label{fig:utility_amazon_pareto}
    \end{subfigure}
    \hfill
    \begin{subfigure}[t]{0.48\textwidth}
        \centering
        \includegraphics[width=\linewidth]{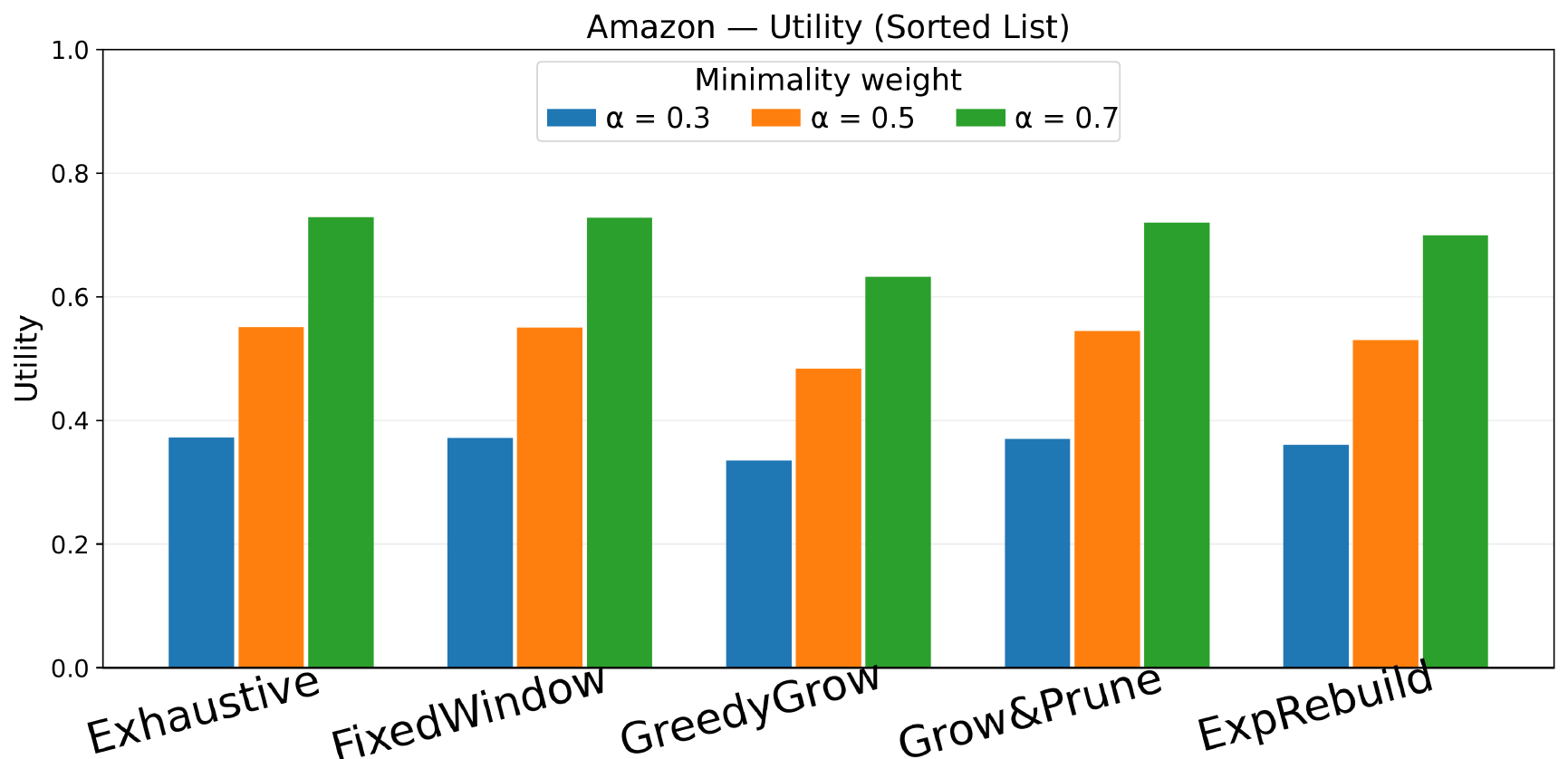}
        \caption{Amazon (Sorted List)}
        \label{fig:utility_amazon_sorted}
    \end{subfigure}

    \caption{Explanation utility for MovieLens and Amazon under varying weights for minimality and interpretability (Group 5-Exst).}
    \label{fig:utility_comparison_10}
\end{figure*}

Figure~\ref{fig:utility_comparison_10} presents the results for groups of size 10. The algorithms show similar trends across utility configurations, suggesting that group size does not substantially affect performance.

\vspace{-4pt}
\stitle{Summary.}
Overall, no heuristic dominates utility across both datasets and all tested weights.
The results show the strengths of each proposed method: ExpRebuild achieves higher utility when interpretability is emphasized, particularly on MovieLens, whereas Grow\&Prune and \mbox{FixedWindow} are more competitive when minimality is given greater weight.

\begin{figure*}
\vspace{-6pt}
    \centering
    \includegraphics[width=0.8\linewidth]{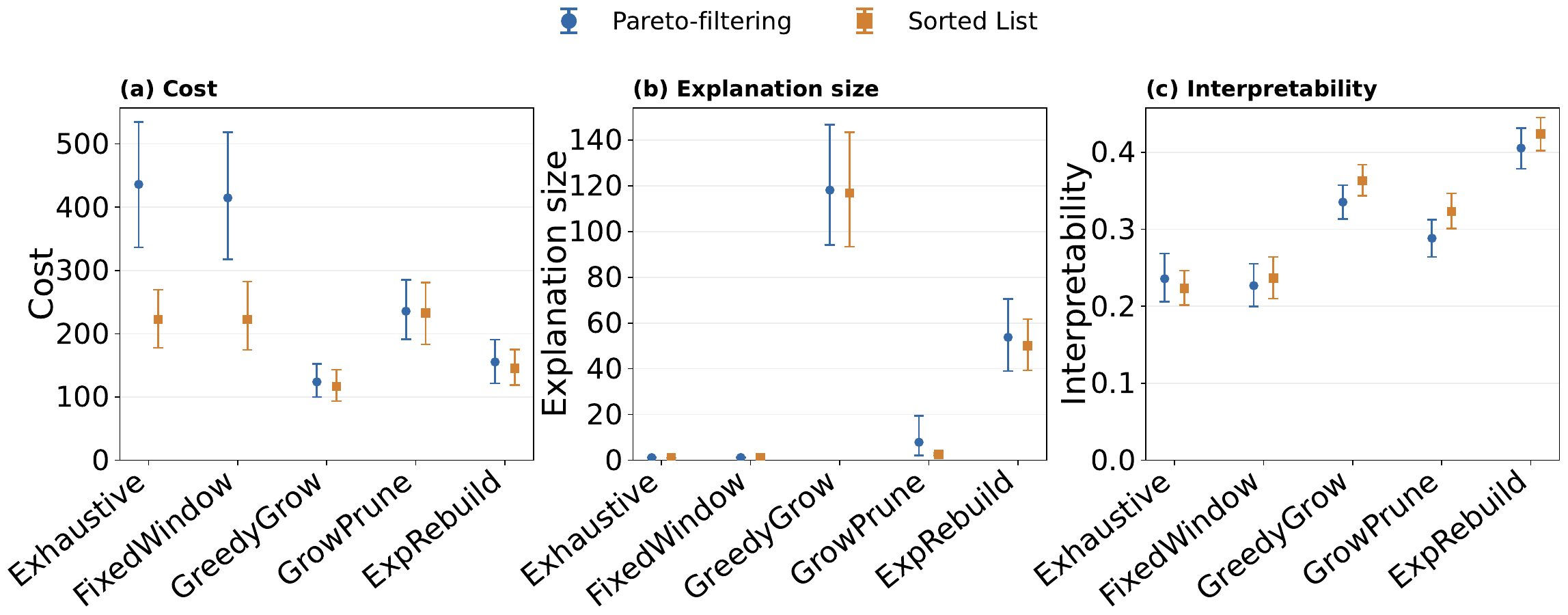}
    \caption{Bootstrap stability analysis on MovieLens for the \mbox{Exhaustive} Search setting ({Group 5-Exst}).
    Points show bootstrap estimates and error bars indicate 95\% confidence intervals from 1,000 group-resampling repetitions for (a)~cost, (b) explanation size, and (c)~interpretability.}
    \label{fig:stability_analysis}
    \vspace{-8pt}
\end{figure*}

\subsection{Stability Analysis}
 We assess the sensitivity of the MovieLens results for groups of five members in the exhaustive search setting (Group 5-Exst).
 For each algorithm and search variant, we resample groups with replacement and recompute the mean over 1,000 bootstrap repetitions, and report the corresponding 95\% confidence intervals. The analysis concerns variation under group resampling rather than independent executions of the algorithms.
Figure~\ref{fig:stability_analysis} shows the uncertainty in (a) cost, (b) explanation size, and (c) interpretability under resampling of the evaluated groups.

 \mbox{GreedyGrow} has the lowest mean cost and ExpRebuild the highest mean interpretability in both search variants.
 With Sorted List, \mbox{GreedyGrow} requires approximately 117 calls, with a 95\% interval of 93--143, compared with 145 calls and an interval of 119--175 for ExpRebuild.
 Their recognition scores are approximately 0.363 and 0.424, with intervals of 0.344--0.384 and 0.402--0.445, respectively.

\mbox{FixedWindow} returns short explanations with Sorted List: approximately 1.14 items, with an interval of \mbox{1.05--1.25}, compared with 1.06 items and 1.02--1.13 for \mbox{Exhaustive}.
Grow\&Prune has a mean size of 2.58 items and an interval of 2.15--3.08 with Sorted List.
Its Pareto-filtering interval is much wider, 2.11--19.46, indicating greater uncertainty in mean explanation size under group resampling.

\vspace{-4pt}
\stitle{Summary.}
Overall, bootstrap resampling preserves the main relative patterns: \mbox{GreedyGrow} has the lowest call cost, ExpRebuild the highest interpretability, and \mbox{FixedWindow} produces short explanations. However, the wider explanation-size interval for Grow\&Prune with Pareto-filtering indicates greater sensitivity to the sampled groups in that configuration.

\subsection{Results Discussion}
\label{sec:discussion}
The evaluation shows that the choice of algorithm depends on the relative importance of explanation size,  cost, and item recognition. Table~\ref{tab:algo_empirical_comparison} summarizes the main findings.
\mbox{GreedyGrow} uses the fewest calls in the evaluated configurations, while \mbox{FixedWindow} returns the shortest explanations among the proposed methods, often at substantially higher call cost.
Grow\&Prune reduces explanation size relative to \mbox{GreedyGrow} through additional refinement, whereas ExpRebuild provides the highest interpretability on MovieLens; on Amazon, the three grow-based methods have similar recognition scores.
\textit{Pareto filtering} offers no uniform improvement and increases \mbox{FixedWindow} cost in every main configuration.

The \textit{exhaustive search} comparison also shows that fewer calls are not an advantage shared by every algorithm: \mbox{FixedWindow} is close to \mbox{Exhaustive} on MovieLens but uses more calls on Amazon.
In the \textit{utility analysis} for groups of 5 members, ExpRebuild leads on MovieLens at weights 0.3 and 0.5 and Grow\&Prune at 0.7, while \mbox{Exhaustive} leads on Amazon with \mbox{FixedWindow} very close.
The \textit{bootstrap analysis} indicates that uncertainty also depends on the variant, particularly for Grow\&Prune explanation size under Pareto filtering.

\section{Related Work}
\label{sec:relatedWork}

Recommender systems are a cornerstone of modern information access, assisting users in navigating large collections of items in domains such as e-commerce, media streaming, and online services. Early approaches were dominated by collaborative filtering and content-based techniques \cite{DBLP:journals/tkde/AdomaviciusT05}, which later evolved into hybrid and deep learning models to improve both scalability and accuracy. Beyond predictive performance, recent research has increasingly emphasized additional dimensions such as fairness, diversity, and explainability \cite{DBLP:journals/vldb/PitouraSK22, DBLP:journals/is/LenziS25, DBLP:journals/ftir/ZhangC20}. We next review existing work on group recommendations and counterfactual explanations. 

\subsection{Group Recommendations} 
Group recommendation has been widely studied~\cite{DBLP:reference/sp/Masthoff15}. A common strategy is to apply a standard recommendation model to each group member and then aggregate the individual recommendation lists~\cite{Amer-Yahia:2009:GRS:1687627.1687713,baltrunas2010group,DBLP:conf/er/NtoutsiSNK12,DBLP:journals/jiis/StratigiPNS22}.

Different approaches have been proposed for the aggregation stage. For instance, \cite{Yuan:2014:CGM:2623330.2623616} considers the varying influence of individual group members when merging recommendations, while \cite{Cao:2018:AGR:3209978.3209998} employs attention networks and neural collaborative filtering to learn aggregation strategies directly from data. Along the same lines, \cite{8731382} combines attention mechanisms with bipartite graph embeddings to capture member influence, whereas \cite{10.1145/2792838.2800190} leverages social networks enriched with user preferences and social interactions to determine group-level recommendations. 

Group members interactions have also been explicitly modeled. For example, \cite{10.1145/3331184.3331251} simulates consensus-building as a series of voting processes and introduces a stacked social self-attention network to learn the underlying voting dynamics. To handle large groups, \cite{8523627} proposes dividing them into subgroups with shared interests, identifying candidate media-user pairs for each subgroup, and then aggregating the resulting collaborative filtering lists. Another strategy is presented in \cite{KIM2010212}, which introduces a two-phase recommender: the first phase seeks to maximize satisfaction at the group level, and the second phase refines recommendations to better satisfy individual preferences. 
 
Utility-based approaches have also been explored. In \cite{Xiao:2017:FGR:3109859.3109887}, each member is assigned a utility score based on the relevance of items, and the system balances these utilities to generate a final group list. Similarly, \cite{10.1145/3297280.3297442} defines user utility as the similarity between individual and group-level recommendations and considers sets of Pareto-optimal items when constructing the list. Finally, \cite{10.1145/3383313.3412232} introduces the notion of rank-sensitive balance, where not only the top recommendation but also each successive prefix of the list must reflect a fair balance of member interests. 

\subsection{Counterfactual Explanations} 
Counterfactual explanations have emerged as a powerful approach to enhance the transparency and reliability of recommender systems. They clarify why certain items are recommended by identifying minimal changes in the input data that would alter the outcome. This form of explanation is intuitive, actionable, and model-agnostic, making it particularly valuable in contexts where user trust and accountability are essential. While counterfactuals have primarily been studied in individual recommendation scenarios, their potential is especially relevant to group settings, where understanding how different members’ preferences shape the joint outcome is critical for transparency and acceptance. 

Recent research has explored counterfactual explanations across diverse recommendation tasks. For instance, CAVIAR~\cite{10.1145/3589335.3651484} modifies image embeddings to reveal the influence of visual features on recommendations, while~\cite{WangLYLX24} introduces CFairER, a fairness-aware counterfactual framework addressing demographic biases such as age and gender. Graph-based methods have also been proposed, including GNNUERS~\cite{10.1145/3655631}, which explains fairness issues in GNN-based recommendation models. Beyond recommended items, \cite{Stratigi2020} studies “why-not” questions in collaborative filtering, providing explanations for items that were absent from recommendations, an idea later extended to graph-based systems in~\cite{10597965}. Temporal dynamics have also been incorporated: the CETD method~\cite{He2023} generates counterfactual explanations for sequential recommendation by considering evolving user behavior. 

Other works focus on general frameworks and model-agnostic solutions. A study in~\cite{10.1145/3459637.3482420} proposes a framework for producing explainable counterfactual recommendations by showing how small changes lead to different results, while \cite{Yao2022LearningTC} presents a learning-based approach that adapts explanations to user interaction histories. MACER~\cite{qingxian_wang_2023} leverages reinforcement learning to provide model-agnostic, item-based counterfactual explanations, particularly suitable in settings with multiple aggregation models, such as group recommendations. LXR~\cite{BarkanBGAK24} offers a post-hoc, self-supervised approach to identify critical user interactions for a recommendation, and \cite{WangLYLX24} studies attribute-level counterfactuals in heterogeneous information networks, focusing on item exposure disparities and fairness. 

Guo et al.~\cite{GuoMGIF2024} use a modified group influence function to estimate the effect of removing training interactions on an individual recommendation. Here, the term ``group'' refers to a set of training samples, rather than a group of users receiving a common recommendation. Its search also relies on model derivatives, whereas our method evaluates observable responses of the group recommender. Sequential counterfactuals consider a different intervention space: GECE~\cite{ScarcelliGECE2025} searches over edited individual interaction sequences using genetic operations, and TRACE~\cite{KimTRACE2026} uses query-limited sequence editing for ranking constraints, including removal of an item from the top-$k$ list. Although the latter target-exclusion criterion is related, these methods explain an individual's ordered sequence rather than a group's joint recommendation.

A recent line of research has begun to explicitly consider group counterfactuals in relation to fairness. For example, CounterFair~\cite{DBLP:conf/icdm/KuratomiLTJPLP24} proposes a framework that generates group counterfactuals for bias detection, fairness-aware recourse, and subgroup identification. By balancing recourse across sensitive groups, CounterFair demonstrates how counterfactual reasoning can serve not only as an explanatory tool but also as a means to mitigate algorithmic bias. This direction complements our focus by highlighting the potential of counterfactuals at the group level, though its emphasis lies in bias detection and subgroup discovery rather than in explaining group recommendations. 

Together, these contributions have laid important foundations for counterfactual reasoning in recommender systems. However, most of this research remains centered on individual recommendations. In contrast, group recommendations introduce additional complexity, as they must reconcile diverse user preferences and balance fairness across members. To the best of our knowledge, counterfactual explanations tailored to group settings have not been systematically explored. Addressing this gap, we investigate how counterfactuals can be adapted to group recommendations, providing explanations that foster both trust and satisfaction among all group members.

\section{Conclusions} 
\label{sec:conc}
In this paper, we introduced the first systematic framework for generating counterfactual explanations in group recommender systems. By formalizing group-level counterfactuals, we highlighted how explanations can reveal the influence of individual interactions on collective outcomes, while also addressing important dimensions such as minimality, interpretability, and fairness. To cope with the inherent combinatorial complexity of explanation discovery, we proposed a family of heuristic algorithms, as well as a \textit{Pareto-based filtering} strategy. Our experimental evaluation on the MovieLens and Amazon datasets demonstrated clear trade-offs between efficiency, explanation size, interpretability, and fairness. In particular, \textit{GreedyGrow} offered low computational cost, while \textit{Grow\&Prune} and \textit{FixedWindow} produced compact explanations, and  \mbox{\textit{ExpRebuild}} showed promising interpretability results. The benefits of Pareto-filtering varied across algorithms and experimental settings.

Despite these findings, our evaluation is limited to static groups, a single recommendation approach, and average aggregation. Furthermore, the proposed heuristics do not generally guarantee globally minimal explanations, and their performance may depend on the characteristics of the underlying dataset.

Looking ahead, several research directions remain open. A promising avenue is to extend the framework to dynamic and sequential group recommendations, where both group composition and preferences evolve over time.


\bibliographystyle{cas-model2-names}
\bibliography{bibliography}

\end{document}